\documentclass{aa}  
\usepackage{graphicx}
\usepackage{graphics}
\usepackage[varg]{txfonts}
\usepackage[colorlinks=true,allcolors=blue]{hyperref} 
\usepackage{enumerate}
\usepackage{amsmath}
\usepackage{natbib}
\usepackage{comment}
\usepackage[para]{threeparttable}
\usepackage{xcolor}
\usepackage[rightcaption]{sidecap}
\usepackage{gensymb}

\begin{document}

   \title{DustPedia and Local Volume Legacy samples \\ as benchmarks for dust evolution in galaxies}

   \subtitle{}

   \author{
            Evangelos D. Paspaliaris\inst{1}\and
            Simone Bianchi\inst{1} \and
            Edvige Corbelli\inst{1} \and
            Angelos Nersesian\inst{2,3} \and
            Fr\'ed\'eric Galliano\inst{4} \and
            Viviana Casasola\inst{5} \and
            Francesco Calura\inst{6} \and
            Emmanuel M. Xilouris\inst{7}\and
            Francesca Pozzi\inst{8}\and
            Georgios Magdis\inst{9,10,11} \and
            Vidhi Tailor\inst{5,8}            
          }

   \institute{INAF - Osservatorio Astrofisico di Arcetri, Largo E. Fermi 5, 50125 Florence, Italy\\
   email: \href{mailto:edpaspaliaris@gmail.com}{edpaspaliaris@gmail.com},
   \and
   STAR Institute, Universit\'e de Li\`ege, Quartier Agora, All\'ee du six Aout 19c, B-4000 Liege, Belgium
   \and
   Sterrenkundig Observatorium Universiteit Gent, Krijgslaan 281 S9, B-9000 Gent, Belgium
   \and
   Universit\'e Paris-Saclay, Universit\'e Paris Cit\'e, CEA, CNRS, AIM, 91191 Gif-sur-Yvette, France
   \and 
   INAF - Istituto di Radioastronomia, Via Gobetti 101, 40129 Bologna, Italy
   \and
   INAF - Osservatorio di Astrofisica e Scienza dello Spazio di Bologna, via Gobetti 93/3, 40129 Bologna, Italy
   \and
   National Observatory of Athens, Institute for Astronomy, Astrophysics, Space Applications and Remote Sensing, Ioannou Metaxa and Vasileos Pavlou, GR-15236 Athens, Greece
   \and 
   Dipartimento di Fisica e Astronomia, Alma Mater Studiorum Università di Bologna, Via Piero Gobetti 93/2, I-40129 Bologna, Italy
   \and
   Cosmic Dawn Center (DAWN), Denmark
   \and
   DTU-Space, Technical University of Denmark, Elektrovej 327, 2800 Kgs. Lyngby, Denmark
   \and
   Niels Bohr Institute, University of Copenhagen, Jagtvej 128, DK-2200 Copenhagen, Denmark
   }
   
   \date{Received -- / Accepted --}

  \abstract
   {}
   {DustPedia and Local Volume Legacy (LVL) are two samples representative of the local galaxy population, including in total $\sim$1000 unique objects of all morphological types, with a wide range of stellar masses and star-formation activity, and a spectral coverage from the ultraviolet to the far-infrared. The purpose of this work is to show 
   that these  samples cover two complementary ranges in stellar mass and galaxy morphology, making them an ideal set for constraining the dominant processes in the evolution of the galactic dust content.}
   {Using the multi-wavelength data provided by the two surveys,
   we fitted the galaxies' spectral energy distribution and estimated their physical properties,  in particular the
   stellar mass, $M_*$, the specific dust mass, $sM_\mathrm{dust}=M_\mathrm{dust}/M_*$, and the specific star-formation rate, sSFR = SFR$/M_*$.} 
   {By combining DustPedia and LVL, we highlight that the trend of $\log_{10}(sM_{\rm dust})$ with $\log_{10}(M_*)$ is not monotonic. Thanks to a large number of objects across a wide range of $M_*$, we have been able to fit two smoothly-joined linear correlations: a positive one for 
   $\log_{10}(M_{*}/$M$_\odot)\lesssim 9.5$ (a range populated mostly by LVL late spirals and irregulars), and a negative one for larger-mass, mainly DustPedia, spirals (with early type galaxies being distinct and more dispersed in the same mass regime). For $\log_{10}(M_{*}/$M$_\odot)> 9.5$, we confirm a strong correlation between $sM_{\rm dust}$ and sSFR; dwarf galaxies, instead, lie below this trend, showing a large scatter of $sM_{\rm dust}$ for -10.5<{$\log_{10}$(sSFR/yr$^{-1}$)}<-9.0.
   By using chemical evolution models we find that the observed $\log_{10}(sM_{\rm dust})$--$\log_{10}(M_{*})$ and $\log_{10}(sM_{\rm dust})$--$\log_{10}$(sSFR) trends can be interpreted mainly by variations in the initial gas mass budget and the galaxy ages, respectively. Low-mass Sm-Irr galaxies with low $sM_{\rm dust}$ and high sSFR can only be reproduced by the models by assuming a highly efficient photofragmentation rate of large grains, and/or low grain-growth in clouds.}
   {}

   \keywords{galaxies: evolution -
   ISM: dust -
   galaxies: ISM -
   galaxies: general -
   galaxies: fundamental parameters -
   galaxies: star-formation}

   \maketitle
%

\section{Introduction}

Metals are produced in stars and are injected into the interstellar medium (ISM) via stellar winds and supernovae (SN) as galaxies evolve \citep{Tinsley1980, MaiolinoMannucci2019}.
A fraction of these metals is condensed into solid grains, during the later stages of stellar evolution [i.e. in SN ejecta and the atmospheres of asymptotic giant branch (AGB) stars;  \citealt{Dwek1998, Calura2025}]. Once in the ISM, metals can accrete into these initial solid seeds and form the bulk of the dust mass \citep{Draine2009, Asano2013, Zhukovska2014}. Dust shields molecules against dissociating radiation and contributes to gas cooling \citep{HollenbachTielens1999, Draine2011}. The chemical composition of the ISM is strongly influenced by the presence of dust, since dust grains provide a surface for the molecules to form and react \citep{SavageSembach1996, Jenkins2009}. 
In other words, the dust content is intrinsically linked to the star-formation history (SFH) of a galaxy. 

Thanks to several space-born and ground-based facilities, such as the {\em Herschel} Space Observatory \citep{Pilbratt2010}, it has been possible to study the spectral energy distributions (SED) of galaxies around the peak of thermal dust emission, from the mid- to the far-infrared (FIR) and sub-millimetre (submm), allowing to trace the bulk of the dust mass,  
$M_{\rm dust}$. 
Using {\em Herschel} data, scaling relations have been studied for the specific dust mass ($sM_{\rm dust}=M_{\rm dust}/M_{*}$, i.e. the ratio between dust and stellar masses), for a few samples of galaxies in the local universe, such as:
the {\em Herschel} Reference Survey \citep[HRS; ][]{Boselli2010}, a volume- and flux-limited sample of 322 galaxies; the 
DustPedia sample \citep{Davies2017, Clark2018},
including 875 galaxies, almost all the largest ($D_{25}>1'$) and nearest ($z<0.01$) galaxies observed by {\em Herschel}; the 
JCMT dust and gas in Nearby Galaxies Legacy Exploration \citep[JINGLE;][]{Saintonge2018}, including 193 galaxies 
with $M_*>10^9 M_\odot$
at $0.01<z<0.05$, 
with data from {\em Herschel} supplemented by submm observations with the James Clerk Maxwell Telescope.
The analysis of these samples has shown that $sM_{\rm dust}$ generally correlates with proxies of the galaxy evolution, decreasing as a function of $M_*$ or gas fraction; and increasing with specific star-formation rate, sSFR=SFR$/M_*$ \citep{Cortese2012, Calura2017, DeLooze2020, Casasola2020}. In particular, the strong correlation between $sM_\mathrm{dust}$ and sSFR confirmed previous results obtained using data from the InfraRed Astronomical Satellite  \citep[IRAS;][]{Neugebauer1984} for a sample of $\sim 1700$ low-redshift galaxies \citep{daCunha2010}.

However, samples such as HRS, JINGLE and DustPedia mainly contain  galaxies
with large stellar masses ($\gtrsim10^9M_\odot$; i.e.\ in later evolutionary stages). Dedicated {\em Herschel} surveys not biased against low $M_*$ values have also been conducted, by choosing low-metallicity objects \citep{RemyRuyer2013} or selecting galaxies by their gas content \citep{DeVis2017_464}; these samples, however, contain a limited number of objects, typically $\lesssim 30$ galaxies with $\log_{10}(M_{*}/{\rm M_{\odot}})<9.0$. Galaxies of such low stellar mass do not show the same scaling laws for $sM_\mathrm{dust}$ detected at high $M_*$. Specifically, $sM_\mathrm{dust}$ increases (rather than decreases) with $M_*$ \citep{DeVis2017_464},
while no correlation is found with sSFR \citep{RemyRuyer2015, DeVis2017_464}. 
By combining these (small) low mass samples, representing the earlier stages of galaxy evolution, to the high mass samples, containing large numbers of more evolved galaxies, attempts have been made to model the evolution of the dust content in galaxies, from the formation of the first grains in the later stages of evolved stars down to dust destruction by SN shocks in the ISM and astration \citep{RemyRuyer2014,RemyRuyer2015, DeVis2017_464, DeLooze2020, DeVis2021, Galliano2021}.

In a recent paper, \citet{Dale2023} examines the scaling laws of the Local Volume Legacy (LVL) survey \citep{Dale2009}, containing 258 galaxies observed by the \textit{Spitzer} space telescope \citep{Werner2004}. For this sample, \citet{Dale2023} find 
that $sM_\mathrm{dust}$ increases with $M_*$, at odds with the results from other large samples such as DustPedia. 
The purpose of this work is to show that the tension between LVL and DustPedia is only apparent, and that the two samples represent two complementary stages in stellar mass, morphology, and thus dust evolution. The paper is structured as follows.
Sect. \ref{sec:sample} describes the data that we have used and the complementarity of the DustPedia and LVL samples. Sect. \ref{sec:CIGALE} presents the spectral energy distribution (SED) fits performed for the two samples. In Section \ref{sec:RatioEvolution} we present and discuss the trends of $sM_{\rm dust}$ with $M_*$ and sSFR, for the combined DustPedia-LVL sample. In Sect. \ref{sec:Models} we compare our results with published results of evolutionary models and results estimated by one-zone dust evolution models. In Sect. \ref{sec:sum} we summarise our findings. 
Finally, in  Appendices~\ref{app:tests}-\ref{app:d20} we provide comparisons of different SED-fitting approaches and further investigate the correlations of the estimated physical properties with morphology.

\section{Samples and data}\label{sec:sample}

We use data available from  the DustPedia project\footnote{\url{https://cdsarc.cds.unistra.fr/viz-bin/cat/J/A+A/609/A37}; \url{https://cdsarc.cds.unistra.fr/viz-bin/cat/J/A+A/624/A80}} and the LVL survey\footnote{\url{https://doi.org/10.26131/IRSA414}}. Both samples consist of local galaxies observed in a wavelength range spanning from the far-ultraviolet (FUV) to the FIR regimes, allowing for an accurate estimation of their fundamental physical properties, such as $M_{\rm *}$,  $M_{\rm dust}$, and SFR. 

\subsection{The DustPedia sample}

The DustPedia sample is comprised of large and nearby galaxies observed by {\em Herschel} \citep[see][for more details]{Davies2017}. Initially, a volume-limited sample of galaxies with $v<3000$~km/s ($\sim$~40~Mpc) was drawn from the HyperLEDA database\footnote{\url{http://atlas.obs-hp.fr/hyperleda/}} \citep{Makarov2014}. Their distribution with morphology is shown in Fig.~\ref{fig:hist} (top panel), where the HyperLEDA Hubble stage $T$ is from a literature compilation of (mostly) visual classifications in the optical. Galaxies were further selected to have large angular sizes, using the HyperLEDA diameter at the B = 25 mag/arcsec$^2$ isophotal level and imposing D$_{25}$ $>$ 1'; and to be detected in the Wide-field Infrared Survey Explorer \citep[WISE;][]{Wright2010} All-sky source catalog\footnote{\url{https://irsa.ipac.caltech.edu/Missions/wise.html}} at a 5-$\sigmaup$ level in the 3.4 $\muup$m band (W1). As shown in Fig.~\ref{fig:hist}, the two criteria (and mostly the size selection) result in a bias against elliptical and later types (in particular irregulars - and unclassified objects), while for types in the range $\-3\leq T\leq7$ about 60\% of the original objects are retrieved, regardless of the detailed morphology. The final DustPedia requirement, the presence of observations in the {\em Herschel} science archive, introduced further biases. While about a quarter of the size-WISE selection is retrieved for $\-4\leq T\leq1$, the fraction of late-type galaxies (LTGs) is smaller ($\sim$15\%  for $2\leq T\leq6$ and $\sim$10\% for $7\leq T\leq10$; see Fig.~\ref{fig:hist}).

The DustPedia datasets for the 875 galaxies in the sample includes photometry executed in a systematic and uniform way across the following bands, GALaxy Evolution eXplorer \citep[GALEX;][]{Morrissey2007}
FUV/NUV; Sloan Digital Sky Survey \citep[SDSS;][]{York2000, Eisenstein2011} ugriz;
2 Micron All-Sky Survey \citep[2MASS;][]{Skrutskie2006}
JHKs; WISE 3.4, 4.6, 12, and 22~$\muup$m; \textit{Spitzer} IRAC (3.6, 4.5, 5.8, and 8.0~$\muup$m) and MIPS (24, 70, 160~$\muup$m); \textit{Herschel} PACS \citep[70, 100, 160~$\muup$m;][]{Poglitsch2010} and SPIRE \citep[250, 350, 500~$\muup$m;][]{Griffin2010}. This process employed aperture-matched techniques, accompanied by comprehensive and compatible uncertainty calculations for all bands. 
Additional photometric data were taken from IRAS (12, 25, 60 and 100~$\muup$m) and from
the {\em Planck} Surveyor
\citep[9 bands including 350, 550 and 850~$\muup$m;][]{Planck2016}.
For a full description of the photometry pipeline we refer the reader to \citet{Clark2018}. 

In this work, we limit the analysis to the objects for which physical properties were derived by \citet{Nersesian2019}. Out of these 814 galaxies, 33\% have an elliptical-lenticular morphology, 54\% are spirals and 13\% are irregulars.

\subsection{The Local Volume Legacy sample}\label{sec:LVL}

The LVL survey provides a sample of 258 galaxies, fully representative of the nearby star-forming population. It consists of all known galaxies within 3.5~Mpc lying outside the Local Group and the Galactic plane ($|b| > 20^\circ$), as well as galaxies in the M81 group and Sculptor filament. These are supplemented by a statistically representative outer tier, derived from the 11 Mpc H$\alphaup$ and Ultraviolet Galaxy Survey \citep{Kennicutt2008} that further consists of two subsets. The primary includes galaxies that meet a combined criterion of $D \leq 11$ Mpc, $|b| > 20^\circ$, $m_{\rm B} < 15$ mag, and $T \geq 0$; the secondary consists of galaxies within 11 Mpc that fall outside one of the limits in Galactic latitude, magnitude, and morphology, but have available H$\alphaup$ data. Out of these, for the LVL outer tier are selected the ones with $|b| > 30^\circ$ and $m_{\rm B} < 15.5$ mag, reaching a completeness of 95\% \citep{Lee2009}. The median distance of the galaxies in LVL is 5.9~Mpc, with the majority of them lying between 0.5~Mpc and 11~Mpc. Dwarf systems constitute 75\% of the sample. LVL survey includes observations from the FUV, up to 160~$\muup$m, i.e. GALEX FUV/NUV, integrated narrow-band H$\alphaup$ line flux corrected for \textsc{Nii} emission by \citet{Kennicutt2008}, $UBVR_CI_C$, SDSS ugriz, 2MASS $JHK_S$, \textit{Spitzer} IRAC (3.6, 4.5, 5.8, 8.0~$\muup$m) and MIPS (24, 70, 160~$\muup$m), and IRAS (12, 25, 60, 100~$\muup$m), if available. The photometry was performed carefully, using the same aperture size across all wavelengths and capturing all the observable emission \citep[for a detailed description of the survey, photometry, and flux properties of the sample see][]{Dale2009}.

Physical properties were derived by \citet{Dale2023} for the 255 LVL galaxies with sufficient photometric coverage; of these, 58 are also included among the DustPedia objects analysed by \citet{Nersesian2019}. Apart from the tests in App.~\ref{app:tests}, we  remove the overlapping objects from LVL and consider them as part of DustPedia only. The independent LVL objects are thus 197, mostly irregulars (61\%) and spirals (31\%). The rest (8\%) are ellipticals and lenticulars.

\begin{figure}[t!]
\centering
\includegraphics[width=0.48\textwidth]{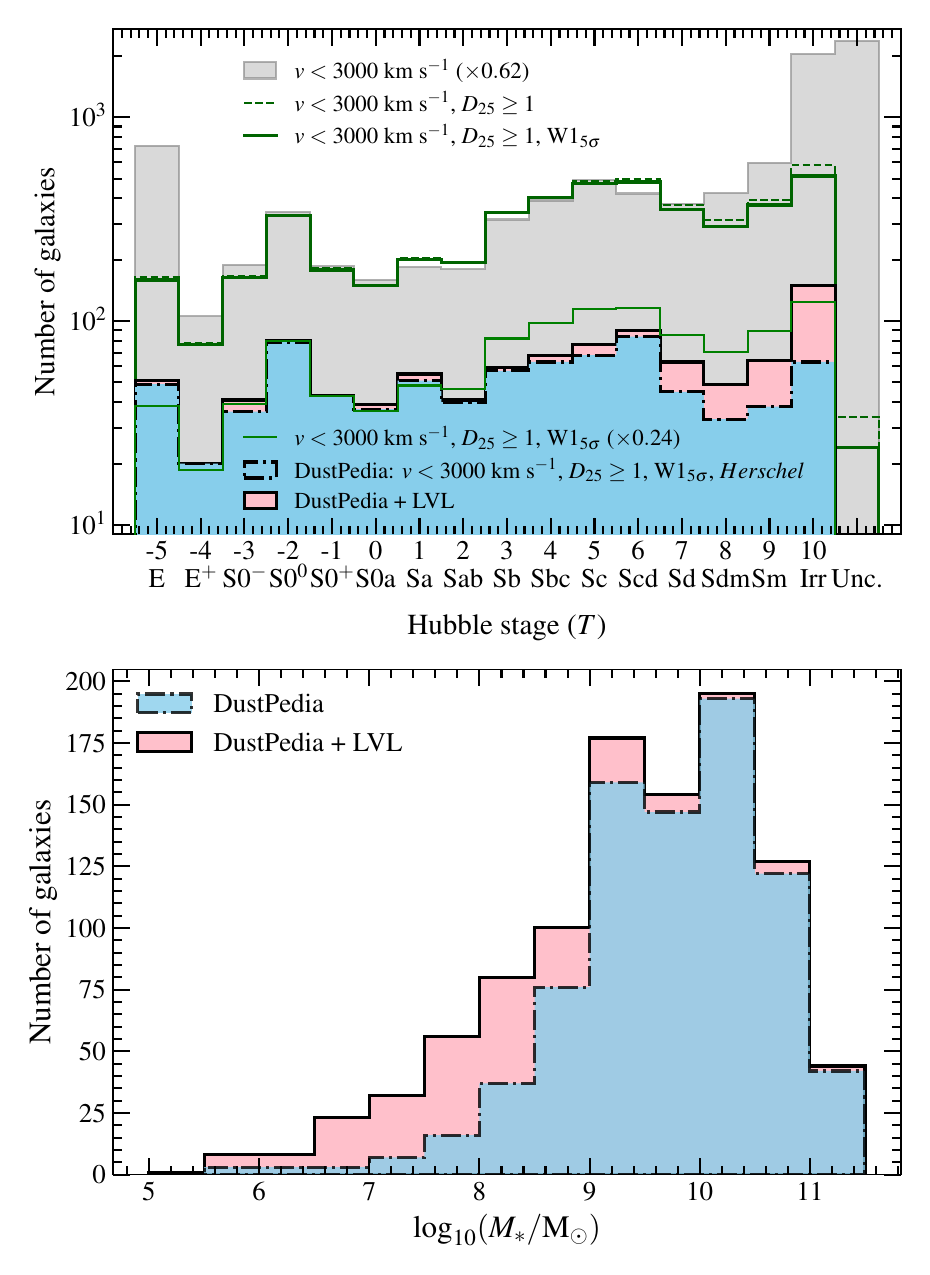}
\caption{Distributions of the morphological type (\textit{top panel}) and stellar mass (\textit{bottom panel}) of the galaxies in our study. Blue histograms represent the distributions of DustPedia galaxies, while the pink ones correspond to the DustPedia + LVL combined sample. In the \textit{top panel} the HyperLEDA parent sample, out of which DustPedia was selected, is plotted with a gray histogram. Changes in the distribution, due to further selection criteria (i.e. $D_{25}>1'$ and W1-band above 5$\sigmaup$) are shown with a dashed and a solid green histogram, respectively. An extra bin includes morphologically unclassified (Unc.) sources.
\label{fig:hist}}
\end{figure}

Because of its morphological composition, LVL complements DustPedia with objects against which the latter was more biased (see Fig.~\ref{fig:hist}, top panel). Anticipating the results of Sect.~\ref{sec:RatioEvolution}, we also show in Fig.~\ref{fig:hist} (bottom panel) that 
LVL is complementary to DustPedia in stellar mass as well.

\subsection{Overlap with other samples}
\label{sec:overlap}

The combined DustPedia-LVL sample, consisting in total of $814+197=1011$ objects, has a significant overlap with many of the {\em Herschel}-based samples with which we will compare our results: 272 galaxies are part of HRS ($\sim$84\% of that sample); 53 are among the 61 objects in the Key Insights on Nearby Galaxies - a Far-Infrared Survey with {\em Herschel} \citep[KINGFISH;][]{Kennicutt2011} project ($\sim$87\% of the sample); 17 are part of the Dwarf Galaxy Sample \citep[DGS;][made of 48 galaxies, thus $\sim$35\% of overlap]{RemyRuyer2013}; 16 are among the 42 galaxies in the {\em Herschel}-ATLAS Phase-1 Limited-Extent Spatial Survey
\citep[HAPLESS;][]{Clark2015}; 16 are part of the $\ion{H}{i}$-selected Galaxies in H-ATLAS sample \citep[H$\textsc{i}$GH;][39\% of its 41 objects]{DeVis2017_464}, 5 of which in their H$\textsc{i}$GH-low subsample for galaxies
with $\log_{10} (M_{*}/$M$_{\odot})<9.0$. Apart from a handful of LVL galaxies, almost all of the overlap is with DustPedia, because of its construction from the post-mission {\em Herschel} Science Archive. \citet{Chastenet2025} recently published another sample based on the {\em Herschel} Science Archive, collecting FIR-submm data for 877 local galaxies within $\sim50$~Mpc belonging to the z = 0 Multiwavelength Galaxy Synthesis project \citep[z0MGS;][]{Leroy2019}; about half of these, 448 galaxies, are in common with our combined sample. There is instead no overlap between DustPedia-LVL and the 193 JINGLE galaxies, because of the mutually exclusive limits in distance selection.

The samples listed in this Section contain only a limited number of objects with $M_*\lesssim 10^9M_\odot$, just as DustPedia alone does. Only LVL provides access to a larger number of low stellar-mass galaxies.

\section{SED fitting}\label{sec:CIGALE}

The multi-wavelength photometry of DustPedia and LVL facilitates the derivation of the galaxies' physical properties, through a SED-fitting analysis:
using this approach, the physical properties of the DustPedia galaxies 
have been derived by \citet{Nersesian2019}, and those of
LVL by \citet{Dale2023}; the \texttt{CIGALE} SED-fitting code \citep{Boquien2019} has been used in both studies. After setting a grid of values for the parameters defining the various modules for the emission by stars, gas, dust, and the dust attenuation, \texttt{CIGALE} generates a library of model-SEDs and through Bayesian inference concludes to the model SED that fits best to the observations. This procedure is performed under the assumption that the energy absorbed and then re-emitted by dust particles is fully conserved. Several modules are available for each component in \texttt{CIGALE} and the estimated physical properties might differ by using different modules or by selecting different values for the free parameters of each module.

The parameter space employed in estimating the physical properties of the DustPedia galaxies is described in detail by \citet{Nersesian2019}. We mention briefly the parameters' setup below. A flexible-delayed star formation history, which allows a late burst or quenching event at 200~Myr before the current moment \citep[module \texttt{`sfhdelayedbq'}; see][]{Ciesla2015} is assumed, with the age of the galaxy varying between 2 and 12 Gyr. The \citet[][BC03]{Bruzual&Charlot} single stellar population module of fixed (solar) metallicity is coupled to the \citet{Salpeter} initial mass function. For the nebular line and continuum emission the default set which is based on \texttt{CLOUDY} templates \citep{Ferland2013, Inoue2011} is used. The stellar and nebular emission are attenuated using the same power-law-modified starburst attenuation curve by \citet{Calzetti2000} and extended by \citet{Leitherer2002} (module \texttt{`dustatt\_calzleit'}). The dust emission modules are estimated by \textit{The Heterogeneous dust Evolution Model for Interstellar Solids} \citep[\texttt{THEMIS};][]{Jones2017}, composed of grains of amorphous (hydro)carbons (a-C, a-C:H) and amorphous silicates with Fe inclusions, whose optical properties have been firmly based on laboratory measurements \citep{Jones2012a, Jones2012b, Jones2012c, Jones2013, Jones2017, Kohler2014}. In our study we have 9 free parameters and a total of about $8\times10^7$ models were produced. The grid of parameters used is listed in Tab.~1 of \citet{Nersesian2019}.

The SED of LVL galaxies have been fitted by \citet{Dale2023} using the same {\tt CIGALE} modules as \citet{Nersesian2019} for SFH, stellar libraries and dust attenuation. However, they used a different IMF \citep{Chabrier2003},  dust emission module \citep[based on the emission templates of][]{Draine2014}, and parameter grid, with 11 free parameters and about $3.4\times10^9$ models. For the sake of uniformity, we repeated the  fitting using the same parameter space and dust model used for DustPedia. Since the LVL data provided by \citet{Dale2023} is not corrected for foreground Galactic extinction, we corrected the fluxes in all bands shorter than $\lambdaup<10~\muup$m, following the methodology presented in \citet{Clark2018}. Upper limits in flux are included in the SED fitting using the default \texttt{'noscaling'} option of {\tt CIGALE}.

\citet{Nersesian2019} find that in 19 DustPedia galaxies the infrared flux might be contaminated by a strong AGN, according to WISE colours and the \citet{Assef2018} 90\%-confidence criterion; however, those objects do not have a SED substantially different from that of galaxies of the same bolometric luminosity \citep{Bianchi2018}. Using {\em Spitzer}-IRAC photometry and the criterion of \citet{DonleyApJ2012}, we find only one LVL galaxy (NGC~5253) that might host a strong AGN. From the \citet{VanVelzen2012} catalogue, \citet{Nersesian2019} selected four more DustPedia objects whose FIR SED might be contaminated by synchrotron and free-free emission from radio lobes (there are no LVL objects matching this selection). Given their limited numbers (23 DustPedia plus 1 LVL galaxies), our results cannot be biased by these objects. Thus, we follow the previous analysis and do not use any {\tt CIGALE} module to fit the contribution of AGNs and radio-lobes; we only mark the objects with different symbols in some of the plots.

From our {\tt CIGALE} fits for LVL galaxies, we use in this work the estimates for $M_*$, $M_\mathrm{dust}$ and SFR (and thus, $sM_\mathrm{dust}$ and sSFR) together with their errors (typically larger for $M_\mathrm{dust}$ and SFR than for $M_*$).In Appendix~\ref{app:tests}, we show that our results are compatible with those presented by \citet{Dale2023}, once the difference in parametrization is taken into account; and that the lack of FIR-submm data for $\lambda>160$~$\muup$m does not bias strongly the results for LVL galaxies.


\begin{figure*}[t!]
\centering
\includegraphics[width=\textwidth]{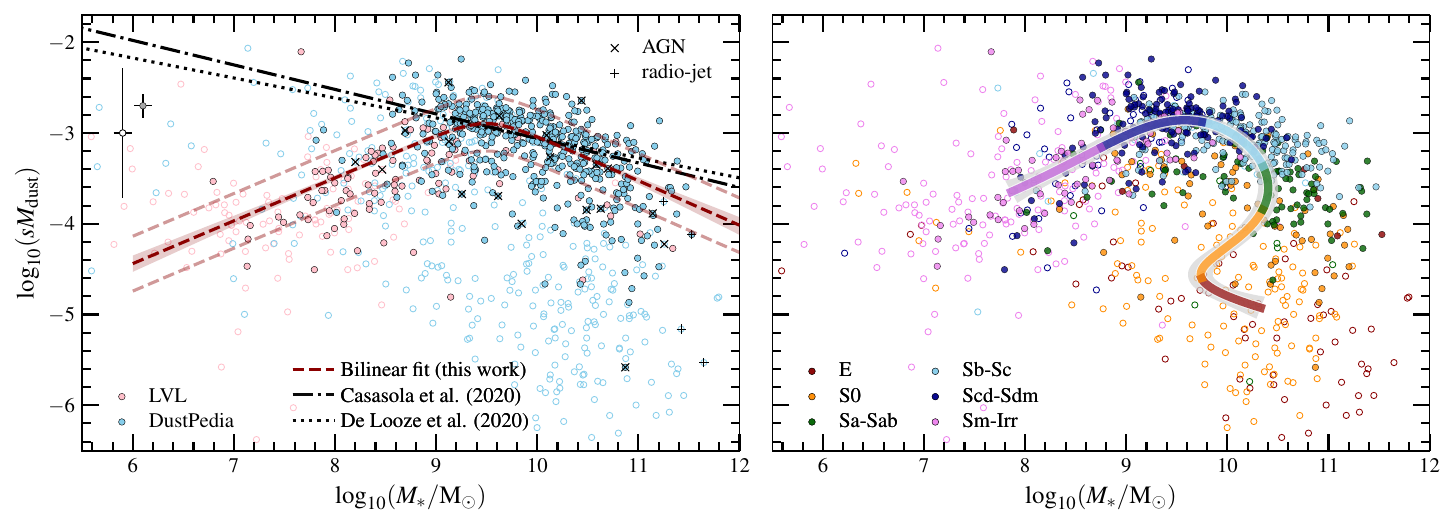}
\caption{$sM_\mathrm{dust}$ as a function of $M_*$. In the \textit{left panel} DustPedia and LVL galaxies are shown in blue and pink, respectively; in the \textit{right} galaxies are colour-coded with morphology. Objects are divided among those with larger (open and paler circles) and smaller uncertainties (filled and darker circles; see text for details): the black error-bars in the \textit{left panel} show the median values of the uncertainties in both ranges. In the \textit{left panel}, the red dashed curve is the smoothly joined bilinear fit for LTGs, the shaded area shows the dispersion around the fit, and the two paler dashed curves indicate the fitted intrinsic scatter. We also show the fits by \citet{Casasola2020} and \citet{DeLooze2020}; and mark galaxies hosting AGNs or radio-jet with a "×" or a "+" symbol, respectively (see Sect.~\ref{sec:CIGALE}). In the \textit{right panel}, the coloured curve shows the combination of the fitted 5$^{\rm th}$ order polynomials, of each physical property, parametrised with Hubble stage (from T=-5 to T=10; pink to dark-red; see Tab.~\ref{tab:recipes}) on the $sM_\mathrm{dust}$--$M_*$ plane. The uncertainty of the polynomials is indicated by the gray shaded area.
\label{fig:R-Mstar-fits}}
\end{figure*}

\section{Scaling laws of specific dust mass}\label{sec:RatioEvolution}

Dust is formed by metals that are synthesised in stars and both the formation and destruction of the grains is connected to the lifecycle of stars. Moreover, it is found that the $sM_{\rm dust}$ can be used as an observational proxy of the molecular gas fraction \citep[see e.g.][]{RemyRuyer2014, Magdis2021} and of the total gas reservoir \citep{Corbelli2012,Orellana2017,Casasola2020,Salvestrini2025,Paspaliaris2025}. Thus, the dust content relates closely to the stellar mass and the star formation activity in galaxies and the $sM_{\rm dust}$ serves as a valuable tool towards understanding the dust production mechanisms and generally the evolution of the ISM in galaxies \citep[see also][]{Calura2017}. In this Section, we show the results of the homogeneous SED fitting of the combined DustPedia-LVL sample. In particular, we present $sM_{\rm dust}$ as a function of $M_{*}$ and of sSFR; 
the latter can be considered also as a proxy of the gas fraction in a galaxy, since the SFR is regulated by the available gas reservoir as indicated by the gas-SFR relation (i.e. the Kennicutt-Schmidt relation; \citealt{Kennicutt1998}; see also \citealt{daCunha2010, Berta2016, DeLooze2020}). Since in general $M_{*}$ increases with a galaxy's age, and the amount of gas relatively to stars decreases (because of astration), the trends of $sM_{\rm dust}$ with $M_{*}$ and (decreasing) sSFR have been interpreted as an evolutionary sequence for the dust content in galaxies \citep{Cortese2012,RemyRuyer2015,Calura2017,DeVis2017_471,DeLooze2020,DeVis2021,Galliano2021}.

\subsection{$sM_{\rm dust}$ vs $M_{*}$}

In Fig.~\ref{fig:R-Mstar-fits} we plot $\log_{10}(sM_{\rm dust})$ as a function of $\log_{10}(M_{*})$. 
For the sake of presentation,
we divide the galaxies according to the uncertainty of the estimates: if an object has 
$\sigma(\log_{10}(sM_{\rm dust}))>0.22$ (i.e. S/N $= sM_{\rm dust}/\sigma(sM_{\rm dust})<2$), it is shown with an open circle; the same criterion is applied to the uncertainty on $M_*$, even though it is $sM_{\rm dust}$ that dominates the selection. The median errorbars of the two uncertainty ranges are shown in the plot. Out of the 1011 objects in the combined sample, the objects with more uncertain estimates are 455: most of them are ellipticals, lenticulars  and irregulars (90.5\%, 69.5\%, and 68.6\%, respectively), with smaller fractions for the other morphological types (15.8\% for Sa-Sab, 8.7\% for Sb-Sc and 28.3\% for Scd-Sdm). Despite the large uncertainties, these estimates are nevertheless useful when trends are studied over several order of magnitudes.
Fig.~\ref{fig:R-Mstar-fits} also highlights with different colors the sample to which a galaxy belongs (either DustPedia or LVL; left panel) and its morphological type (right panel). In order to visualize better the trends with morphology, we have fitted a 5$^{\rm th}$ order polynomial to $M_{*}$ and $sM_{\rm dust}$ as a function of the Hubble stage $T$ (See App.~\ref{app:morpho}); the two fits are combined and plotted in the right panel of Fig.~\ref{fig:R-Mstar-fits}, colour coded by morphological type.

The left panel of Fig.~\ref{fig:R-Mstar-fits} confirms visually what we anticipated in Sec.~\ref{sec:LVL} and Fig.~\ref{fig:hist}. LVL and DustPedia are complementary not only in morphologies, but also in stellar masses. The bulk of the DustPedia galaxies have $8<\log_{10} (M_{*}/$M$_{\odot})<11$, with a median value of 9.75, while the majority of the LVL galaxies have lower stellar masses, with $6.5<\log_{10} (M_{*}/$M$_{\odot})<9.5$ and a median value of 8.19. Fig.~\ref{fig:R-Mstar-fits} also shows that the trends for the low and high $M_{*}$ ranges are different: $sM_{\rm dust}$ increases with $M_{*}$ for lower stellar masses, reaching a peak at $\log_{10}(M_{*}/\rm M_{\odot})\approx9.5$, beyond which the trend is reversed. The galaxies along the increasing trend are mainly Sm-Irr, with fewer Scd-Sdm (the majority being LVL objects), while the rest of the morphological types follows the decreasing trend (and are mostly from DustPedia), with the early-types (E, S0) being distinct from the spirals (Sa-Sdm), and more scattered. The Spearman's correlation coefficients in both mass ranges are very similar (in absolute value): we obtain $\rho_{\rm S}=0.43$ when selecting galaxies with $\log_{10}(M_{*}/{\rm M_{\odot}})<9.5$, $\rho_{\rm S}=-0.45$ for $\log(M_*/M_\odot)>9.5$; they raise to $\rho_{\rm S}=0.57$ and $-0.56$ for the two stellar mass ranges, respectively, when only LTGs ($T>0.5$) are considered.
While the negative correlation for high stellar masses is well established in the literature, as we will discuss later, the positive correlation for the low-mass end echoes that found by \citet{Dale2023} using LVL data only, i.e. the motivation itself for the current work. 
The correlation claimed by those authors $\rho_{\rm S}=0.58$ (without any cut in morphology), is stronger than what is measured here,
probably because of the exclusion of those galaxies for which they only provide upper limits in $M_{\rm dust}$.

We used the {\tt Python UltraNest} package for Bayesian inference \citep{Buchner2021} to fit the datapoints with a function that smoothly joins, at intermediate stellar masses, a negative and a positive linear correlation,  i.e., 
\begin{equation}
\log_{10}(sM_\mathrm{dust})  = 
-\epsilonup \log_{10} \left(  
10^{-\frac{1}{\epsilonup} \log_{10} (sM_\mathrm{dust}^\mathrm{low})}+
10^{-\frac{1}{\epsilonup} \log_{10} (sM_\mathrm{dust}^\mathrm{high})}
\right) 
\end{equation}    
The intrinsic scatter of the datapoints, $\sigmaup$, is also fitted.
The fit considers errors on both axes, for all datapoints, regardless of the error magnitude. When selecting only LTGs, we obtain
\begin{eqnarray*}
\log_{10}\left(sM_\mathrm{dust}^\mathrm{low}\right) &=& \left(0.47^{+0.04}_{-0.04}\right) \times \log_{10} (M_*/M_\odot) + \left(-7.27^{+0.30}_{-0.31}\right),\\
\log_{10}\left(sM_\mathrm{dust}^\mathrm{high}\right) &=& \left(-0.49^{+0.06}_{-0.07}\right) \times \log_{10} (M_*/M_\odot) + \left(1.87^{+0.72}_{-0.59}\right),\\
    \epsilonup &=& 0.33^{+0.19}_{-0.16},\\
    \sigmaup &=& 0.30^{+0.01}_{-0.01}.
\end{eqnarray*}
Results of the fit are shown in the left panel of Fig.~\ref{fig:R-Mstar-fits}, where we plot the median representation of the model,
together with the dispersion given by the 0.16 and 0.84 percentiles around the median, and the intrinsic scatter. While the parameter defining the smoothing, $\epsilon$, is poorly constrained, those for the two linear correlations are very similar to what would have been obtained by fitting separately the data-points below and above $\log (M_*/M_\odot)=9.5$ (which is also close to the position of the maximum of the joining function). When all morphological types are considered, instead, the slope will become flatter for galaxies with lower stellar masses, and steeper (in absolute value) for higher stellar masses (not shown), as a result of the 
lower (and more uncertain) values of $\log_{10}(sM_\mathrm{dust})$ for early-type galaxies (ETGs), and of their general negative correlation with $\log_{10}(M_{*})$.
Correlations and fits for each morphological type are shown in App.~\ref{app:morpho}. 

\begin{figure*}[t!]
\centering
\includegraphics[width=\textwidth]{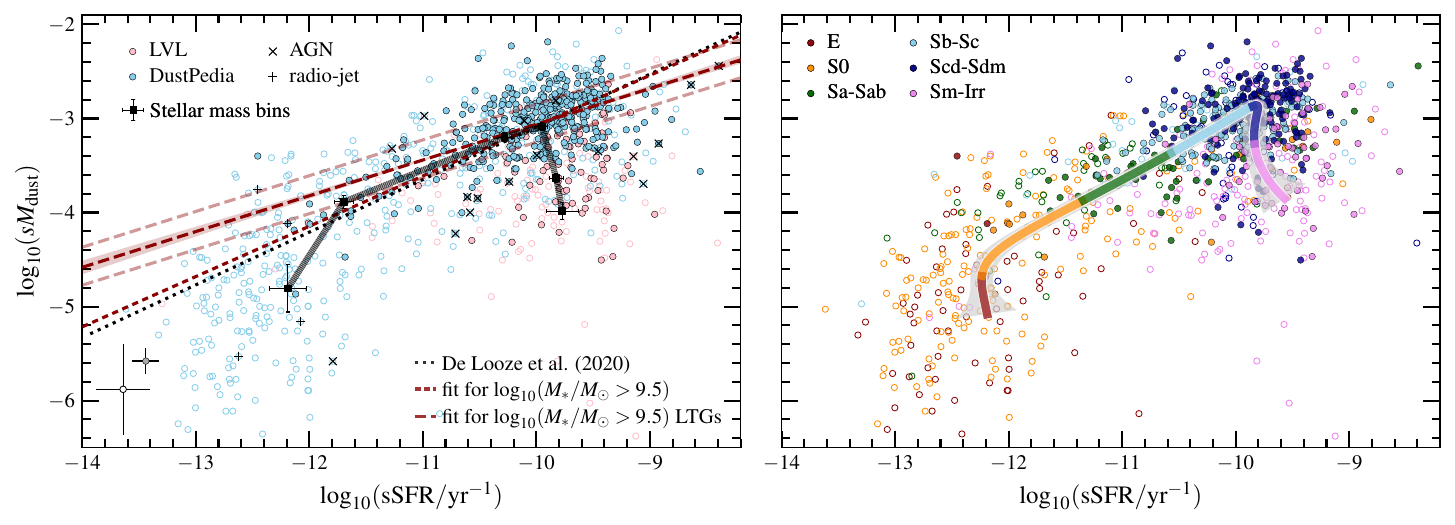}
\caption{$sM_\mathrm{dust}$ as a function of sSFR. Datapoint colors and shapes, median errorbars, and marks indicating AGN and radio-jets (\textit{left panel}) are same as in Fig.~\ref{fig:R-Mstar-fits}.
In the \textit{left panel}, the red dashed curve is the linear fit to LTGs with $\log_{10}(M_*/\mathrm{M}_\odot)>9.5$, the shaded area the $1\sigma$ dispersion around the fit, and the two paler dashed curves the fitted intrinsic scatter 
(the short-dashed line being the fit including also ETGs). The dotted line is the fit by \citet{DeLooze2020}.
The black squares show the median, and standard deviation, for bins in $M_{*}$ of 1.0 dex width, centred from $\log_{10}(M_*/\mathrm{M}_\odot)=7$ (on the right side of the panel) to $\log_{10}(M_*/\mathrm{M}_\odot)=12$ (on the left side). In the \textit{right panel}, the coloured curve shows the combination of the fitted 5$^{\rm th}$ order polynomials, of each physical property, parametrised with Hubble stage (from T=-5 to T=10; pink to dark-red; see Tab.~\ref{tab:recipes}) on the $sM_\mathrm{dust}$--sSFR plane. The uncertainty of the polynomials is indicated by the gray shaded area. 
\label{fig:R-sSFR-fits}}
\end{figure*}

\subsection{$sM_{\rm dust}$ vs \textit{sSFR}}

In Fig.~\ref{fig:R-sSFR-fits} we plot $\log_{10}(sM_{\rm dust})$ as a function of $\log_{10}$(sSFR). In analogy with
Fig.~\ref{fig:R-Mstar-fits}, data-points are colour coded by sample (left panel) and morphology (right panel); open circles are used for galaxies with uncertain estimates. In this plot, also the uncertainty for the quantity on the y-axis, sSFR, can be significant. Thus, defining objects with uncertain estimates those with $S/N$<2 for $sM_{\rm dust}$ or sSFR results in a larger selection. These galaxies are now 557 (97.3\% among all E, 85.7\% of S0, 46.3\% of Sa-Sab, 20.8\% of Sb-Sc, 31.2\% of Scd-Sdm, and 70.0\% of Sm-Irr galaxies). Also, galaxies that were considered as having more certain estimates in Fig.~\ref{fig:R-Mstar-fits}, because they have $S/N$>2 for $sM_{\rm dust}$, could be classified as uncertain in Fig.~\ref{fig:R-sSFR-fits} because they have $S/N$<2 for sSFR; this is reflected in the different median errorbars in Figs.~\ref{fig:R-Mstar-fits} and~\ref{fig:R-sSFR-fits}.

Fig.~\ref{fig:R-sSFR-fits} shows that $\log_{10}(sM_{\rm dust})$ in general increases with increasing sSFR, with most of the objects from DustPedia following the increasing trend, while  those of LVL are mostly below it, dragging the trend at lower $sM_\mathrm{dust}$ for high sSFR. Indeed, most LVL galaxies have similar $\log_{10}$(sSFR/yr$^{-1}$) values, between -10.2 and -9.2, but a wide spread of $\log_{10}(sM_{\rm dust})$ values, from -2.8 to -4.5. This trend becomes evident when we plot the running median of all DustPedia-LVL galaxies in bins of $\log_{10}(M_*/{\rm M}_\odot)$. 
Galaxies of lower $M_{*}$ have higher sSFR. As we have also seen in Fig.~\ref{fig:R-Mstar-fits}, while $M_{*}$ increases, $sM_\mathrm{dust}$ increases until it reaches a peak, at $\log_{10}$(sSFR/yr$^{-1})\simeq-9.9$; this happens with a small variation ($<0.5~\rm{dex})$ in $\log_{10}$(sSFR). Going towards the highest $M_*$ bins, both the $sM_{\rm dust}$ and the sSFR decrease.
The right panel of Fig.~\ref{fig:R-sSFR-fits} shows that the place where galaxies lie in the diagram also depends on their morphology. For instance, although Sm-Irr galaxies have similar sSFR with the late spirals (Scd-Sdm), they have lower $sM_{\rm dust}$, in accordance to their lower $M_{*}$. Similarly to the trend suggested by the stellar mass bins, the combination of the fitted 5$^{\rm th}$ order polynomials of the plotted properties (see App.~\ref{app:morpho}), shows the $sM_{\rm dust}$ rapidly increasing with relatively constant sSFR for the Sm-Irr and Scd-Sdm bins. In this case the peak $sM_{\rm dust}$ occurs at a slightly higher $\log_{10}$(sSFR/yr$^{-1} (= -9.8)$. The $sM_{\rm dust}$ then decreases with decreasing sSFR, for earlier stage galaxies. 

The correlation between $\log_{10}(sM_{\rm dust})$ and $\log_{10}$(sSFR) for all objects has Spearman's coefficient, $\rho_{\rm S}=0.62$. Because of what we discussed in the previous paragraph, the significance of the correlation increases when the sample is restricted to galaxies with $\log_{10}(M_*/{\rm M}_\odot) > 9.5$: in this case we have $\rho_{\rm S}=0.85$ (for 525 objects; equivalent results can be obtained by selecting DustPedia galaxies only). A fit to these objects (red line, small dashes; Fig.~\ref{fig:R-sSFR-fits}, left panel) yields
\begin{equation}
\log_{10}(sM_{\rm dust}) = 
\left(0.53\pm0.02\right)\times\log_{10}({\rm sSFR}/{\rm yr}^{-1}) + \left(2.3\pm0.2\right),
\end{equation}
with an intrinsic scatter $\sigmaup=0.27\pm0.01$ dex. If we restrict the fit to the massive LTGs ($T>0.5$, 321 objects) only (red line; long dashes), we get
\begin{equation}
\log_{10}(sM_{\rm dust}) = 
\left(0.38\pm0.02\right)\times\log_{10}({\rm sSFR}/{\rm yr}^{-1}) + \left(0.74\pm0.21\right),
\end{equation}
with a reduced intrinsic scatter, $\sigmaup=0.20\pm0.01$ dex; the correlation coefficient, however, is smaller, $\rho_{\rm S}=0.65$, because of the more restricted dynamic range of $\log_{10}({\rm sSFR})$.
When fits are done for individual morphology bins (App.~\ref{app:morpho}) a positive trend is found in all cases, again with different normalizations and intrinsic scatter: the smallest scatter is for Sa-Sab and Sb-Sc; it becomes larger for Sm-Irr, whose fit is below that of the later type spirals, in accordance to what discussed so far; it is largest for ellipticals.

\subsection{Comparison with previous works}

The negative correlation between $\log_{10}(sM_{\rm dust})$ and $\log_{10}(M_{*})$ at higher stellar masses, shown in Fig.~\ref{fig:R-Mstar-fits}, is a well known result in the literature and has been found for a wide diversity of samples of galaxies in the local Universe. For instance, it has been described for HRS \citep{Cortese2012}; for the {\em Planck}-selected sample of \citet{Clemens2013}; for the JINGLE sample, united with HRS, KINGFISH, HAPLESS and H$\textsc{i}$GH \citep{DeLooze2020}; and for the LTGs in DustPedia \citep{Casasola2020}.
Only \citet{Chastenet2025} find no correlation for the z0MGS galaxies they study but, as they admit, this is mainly due to the exclusion of objects with low sSFR from the analysis.
The normalization and the slope of the correlation are found to vary depending on various selections: for instance, \citet{Cortese2012} and \citet{DeLooze2020} find that $sM_{\rm dust}$ is systematically lower for HI-deficient galaxies, while \citet{Orellana2017} - using a sample of 1630 galaxies with {\em Planck} detection - 
derived a steeper correlation for starbursts than for more quiescent ones. The inverse correlation was confirmed by \citet{Donevski2020} on a  more distant sample (up to z$\simeq$5), but with a median $sM_{\rm dust}$ more than an order of magnitude higher than local early- and late-types, indicating an evolution with redshift. Two fits from the literature are shown in Fig.~\ref{fig:R-Mstar-fits}; that of
\citet{Casasola2020}, obtained for DustPedia LTGs with the same {\tt CIGALE} dataset used in this work; that of \citet{DeLooze2020} for their 
JINGLE-based sample (corrected for their different modelling parametrization; see Appendix~\ref{app:d20}). The two fits are fully compatible between themselves. Since the range of their analyses extends to lower $M_{*}$, their fitted linear trends are flatter than what is obtained here for high $M_{*}$ only, but overall consistent with the current results. 

Instead, the positive trend of $sM_{\rm dust}$ vs $M_{*}$, for low $M_{*}$, is described only in a few literature works and required carefully crafted samples, because of the difficulty of estimating the properties for low-mass galaxies. \citet{Grossi2015} studied the star-forming dwarfs in the {\em Herschel} Virgo Cluster Survey \citep[HeViCS;][]{Auld2013}, combined with similar objects from KINGFISH and DGS, and found hints of a change in trend with respect to more massive KINGFISH and HeViCS galaxies, but with a large scatter that prevented the detection of any correlation. 
\citet{DeVis2017_464}, studying 17 galaxies with 
$\log_{10} (M_{*}/$M$_{\odot})<9$ (the H$\textsc{i}$GH-low sample), find a different trend with respect to more massive objects. 
Suggestions for a change in slope at small $M_{*}$ can be found also in other samples (see, for example, Fig.~11 in \citealt{Casasola2020} and Fig.~2 in \citealt{DeLooze2020} for $\log_{10}(M_{*}/{\rm M_{\odot}})<9$), again for a limited number of objects. 

The strong correlation between $\log_{10}(sM_{\rm dust})$ and $\log_{10}$(sSFR) shown in Fig.~\ref{fig:R-sSFR-fits} was already found by \citet{daCunha2010}, with $\rho_{\rm S}=0.84$ for a sample of $\sim3200$ galaxies; they also demonstrated that this is not driven by the $M_{*}$ (used at the denominator of both the $sM_{\rm dust}$ and sSFR). The correlation has been confirmed for several other {\em Herschel}-based samples: for HRS, by \citet{Cortese2012}, using NUV-$r$ as proxy for sSFR; for KINGFISH \citep{RemyRuyer2015}; for HRS, HAPLESS and H\textsc{i}GH  \citep{DeVis2017_464}; for those samples above, plus JINGLE \citep{DeLooze2020}; for DustPedia and the DGS \citep{Galliano2021}. 
The fit by \citet{DeLooze2020}, including galaxies of all morphological types, is shown in the left panel of Fig~\ref{fig:R-sSFR-fits} (after the corrections described in Appendix~\ref{app:d20}); it is analogous to what we find in this work without a morphological selection. While the paucity of objects in the local Universe does not allow to extend the analysis at the $\log_{10}$(sSFR/yr$^{-1})>-9.4$, we note here that samples of high redshift galaxies have shown that the general correlation of $\log_{10}(sM_{\rm dust})$ versus $\log_{10}$(sSFR) extends also beyond that limit \citep{Rowlands2014,DeVis2017_464,Donevski2020}.

\begin{figure*}[t!]
\centering
\includegraphics[width=\textwidth]{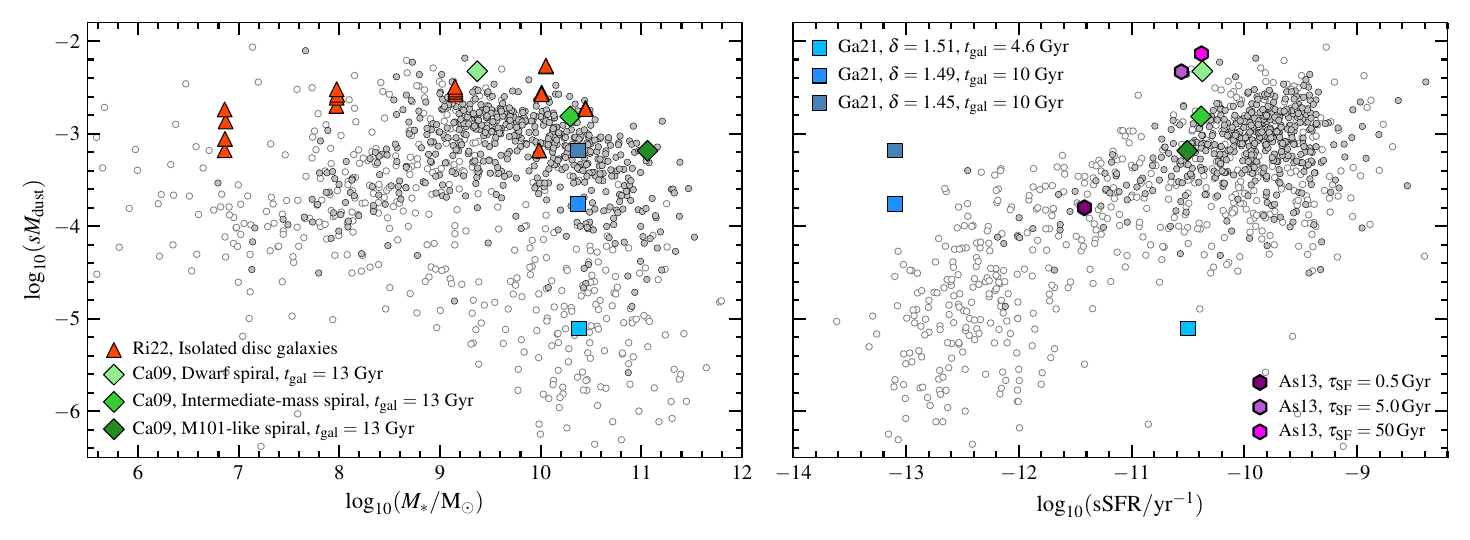}
\caption{$sM_\mathrm{dust}$ as a function of $M_{*}$ (\textit{left panel}) and sSFR (\textit{right panel}), compared to model predictions from the literature. Galaxies with uncertain estimates are shown with  open symbols. Model estimates by \citet[Ga21]{Galliano2021} are shown with blue squares; lighter blue colour correspond to higher outflow rate ($\delta_{\rm out}$). The models for three spiral galaxies of different baryonic masses from \citet[Ca09]{Calura2009} are plotted with green diamonds, with darker green indicating higher assumed baryonic mass. Predictions for isolated disc galaxies by the \citet[][Ri22]{Richings2022} simulations are also plotted in the \textit{left panel}. In the \textit{right panel}, together with the Ca09 and Ga21 estimations, are also plotted the ones of \citet[As13]{Asano2013} with violet hexagons; lighter colours represent higher star-formation timescales ($\tauup_{\rm SF}$). Predictions by Ri22 for the right panel, and by As13 for the left panel, are not available in the literature.
\label{fig:models_literature}}
\end{figure*}

Also for $\log_{10}(sM_{\rm dust})$ versus $\log_{10}(sSFR)$, the departure of galaxies of lower stellar mass from the main scaling law of Fig.~\ref{fig:R-sSFR-fits} has been found and described only for a few samples of galaxies with specific properties. For similar values of sSFR, \citet{DeVis2017_464} find lower $sM_\mathrm{dust}$ than the main trend for the HiGH-low galaxies. Another sample showing no significant correlation between $sM_\mathrm{dust}$ and sSFR is the DGS, made by 48 star-forming and low-metallicity dwarfs, which have lower $sM_\mathrm{dust}$ 
and higher sSFR than expected from the trend shown by KINGFISH galaxies \citep[see Fig.~11 in][]{RemyRuyer2015}.

\subsection{Advantages and caveats}

We have shown in this Section that the combination of LVL and DustPedia allows to study the main scaling laws of $sM_{\rm dust}$, and the departures from them for different galaxy subsets, in more details than what was previously possible by including a small number of low-mass objects to larger samples of more massive galaxies. For example, \citet{DeVis2017_464} did not attempt to fit the $sM_{\rm dust}$ vs $M_{*}$ trend because of the small number of objects in their H$\textsc{i}$GH-low sample of 17 galaxies with $\log_{10}(M_{*}/{\rm M_{\odot}})<9$. 
DustPedia-LVL, instead, contains 309 galaxies in the same $M_{*}$ range; thus, we could fit also the low mass regime. Similarly, \citet{Galliano2021} studied the evolution of dust in galaxies by supplementing DustPedia with 34 DGS objects, while LVL adds 197 objects to the larger sample.

However, a caveat for the use of LVL should be mentioned, i.e. the lack of FIR/submm data for $\lambda>$ 160~$\muup$m. In Appendix~\ref{app:tests}, we discuss the possibility of an underestimate of  $M_{\rm dust}$ (and thus of $sM_{\rm dust}$): if we correct the dust masses of LVL galaxies according to the (rather uncertain) fit shown in Fig.~\ref{app:fig-comp-LVL-DP}, low-$M_*$ galaxies will show a shallower trend of $\log_{10}(sM_{\rm dust})$ versus $\log_{10}(M_*)$ than that presented in Fig.~\ref{fig:R-Mstar-fits}, with Spearman's $\rho_{\rm S}=0.3$; yet the dichotomy with the trend of high $M_\star$ galaxies (unaltered by the correction) will remain. Similarly, the difference between $sM_{\rm dust}$ for high- and low-$M_*$ galaxies at high sSFR will remain, though much reduced. Currently, LVL galaxies with $\log_{10}($sSFR$/\mathrm{yr}^{-1})\approx -10$ and $\log_{10}(M_{*}/M_\odot)\approx 8$ have $sM_{\rm dust}$ values 0.8 dex lower than their $\log_{10}(M_{*}/M_\odot)>9.5$ counterparts; this will reduce to just 0.2~dex after the correction. Future FIR/submm observations of LVL galaxies beyond the peak of thermal emission are needed to settle this issue.


\section{Dust evolution modelling}\label{sec:Models}

Several works have modelled the variation of the dust content (and other integrated properties) of galaxies as a function of cosmic time \citep[e.g.][]{DwekScalo1980, Dwek1998, LisenfeldFerrara1998, Hirashita1999, MorganEdmunds2003, Inoue2003, Dwek2007, Galliano2008, Calura2008, Valiente2009, MattssonAndersen2012, Asano2013, Zhukovska2014, Feldmann2015, DeLooze2020, Nanni2020, DeVis2021, Sawant2025}. In particular, limited samples of galaxies of low stellar mass (i.e. H\textsc{i}GH and DGS) have been used to define the earlier stages of dust evolution \citep{DeVis2017_471,Galliano2021}. 

In this Section, we explore the impact of the DustPedia-LVL dataset on dust modelling, and in particular of the estimates for the larger number of low-stellar mass galaxies available from LVL. We first compare the scaling laws presented in the previous Section with published evolutionary models; we then use the \texttt{chemevol}\footnote{\url{https://github.com/zemogle/chemevol}} \citep{DeVis2021} to understand which parameters for dust evolution can better describe the dataset presented here.

\subsection{Models from literature}

We first compare the DustPedia-LVL results with four models for $sM_\mathrm{dust}$ we selected from the literature:
\citet[Ca09]{Calura2009}, whose estimations for $sM_\mathrm{dust}$, $M_\mathrm{*}$, and sSFR for spiral galaxies of three different baryonic masses $(M_\mathrm{bar})$ are taken from \citet{Calura2017}; \citet[As13]{Asano2013}, whose estimations for $sM_\mathrm{dust}$ vs sSFR, for a total baryon mass (stars+ISM) of $10^{10}\rm~M_\odot$ and three different star-formation timescales ($\tauup_{\rm SF}$; i.e. the ISM depletion timescale) are taken from \citet[the trend vs $M_*$ is not available]{RemyRuyer2015}; the models of \citet[Ga21]{Galliano2021} for an initial gas mass of $4\times10^{10}\rm~M_\odot$ and three different outflow rates, presented in their Fig.~15; and the results 
from hydrodynamical simulations of isolated disc galaxies by \citet[Ri22; only $sM_\mathrm{dust}$ vs $M_*$ is presented by the authors]{Richings2022}, with $M_{*,\rm ~tot}$ ranging from $6.6 \times 10^{6} \rm ~M_\odot$ (dwarfs) to $3.1 \times 10^{12} \rm ~M_\odot$ (massive galaxies), in five steps, with two additional cases where the gas fraction is increased (or decreased) by 20\%, for $M_{*,\rm ~tot}=1.1\times 10^{10}\rm ~M_\odot$.

The evolution frameworks developed by Ca09 and As13 incorporate the same fundamental physical mechanisms: dust formation in the ejecta of core-collapse SN and AGB stars, subsequent grain growth in the dense ISM, and dust removal through SN-driven destruction and astration. Their treatments diverge primarily in the environmental regulation of these processes. In Ca09 model, the spiral galaxy is dominated by a thin disc of stars and gas, consisting of several independent rings without exchange of matter among them. They adopted an inside out disc formation, with the timescale for disc formation increasing with galactocentric distance, and assume that the star-formation efficiency (SFE) is higher for more massive objects. As13 employed a closed-box model that isolates the intrinsic interplay between stellar dust production and accretion, allowing them to identify the critical metallicity, above which grain growth becomes the dominant channel. Ga21 rewrote the \texttt{chemevol} version of \citet{DeVis2017_471}, which, adopts the same physical mechanisms, as the two aforementioned models (i.e. Ca09, As13), for dust production and destruction, but embed these processes within a more dynamically realistic context that includes gas inflows, outflows, and galactic fountains (i.e. recycling of the outflows), each of which can dilute, expel, or destroy dust and thereby modulate the efficiency and timing of the underlying mechanisms. In their version, Ga21 extended the code's formalism by applying a hierarchical Bayesian framework to constrain the efficiencies of SN dust condensation, grain growth, and shock destruction across a large sample of galaxies. The Ri22 models are computed with the hydrodynamics code \textsc{gizmo} \citep{Hopkins2015} assuming a rotating disc of gas and stars, along with a spherical central stellar bulge, all embedded within a live dark matter halo. The parameters of the models (e.g. bulge-to-total ratio, half-light radius, gas fraction, and metallicity) are accordingly chosen in order to follow the scaling relations of nearby typical spiral galaxies at redshift zero. Within the simulations other processes are implemented like star formation and feedback by SN, stellar winds, photoionization of the surrounding gas, and stellar radiation pressure, as well as non-equilibrium chemistry covering a wide range of gas phases, from cold dense molecular clouds, to hot highly ionised plasma. 

The comparison between DustPedia-LVL and the literature models is shown in Fig.~\ref{fig:models_literature}. We note that, while we fitted galactic properties adopting the \citet{Salpeter} IMF, this is used by Ga21 only; instead, Ca09 adopt \citet{Scalo1986}, As13 \citet{LarsonIMF}, and Ri22 \citet{Kroupa2001}.
Unfortunately, conversion factors are not available in the literature between \citet{Salpeter}, typically producing higher values for $M_*$ and SFR, and all the other IMFs above: if we had applied the conversion from the \citet{Kroupa2001} IMF (as in \citealt{2014ARA&A..52..415M}), and if conversion from other IMFs are similar, the modelled $sM_\mathrm{dust}$ would have been about 0.2~dex lower, $M_\mathrm{*}$ 0.2 dex higher (and sSFR almost unchanged), i.e. by about a symbol size for models in Fig.~\ref{fig:models_literature}. Furthermore, other different assumptions on galactic modelling might have resulted in other shifts (see App.~\ref{app:tests} and~\ref{app:d20}). Thus, we apply no correction but keep in mind that possible offsets between models and between a model and our results might be due to this issue. Looking at the $\log_{10}(sM_{\rm dust})$--$\log_{10}(M_{*})$ plane (left panel of Fig.~\ref{fig:models_literature}), the Ca09 model follows the negative trend for higher $M_*$, found for our data (apart from a systematic offset towards higher $sM_{\rm dust}$ values). The Ga21 models show that for the same $M_*$, galaxies exhibiting stronger outflows ultimately display lower, $sM_{\rm dust}$, providing an interpretation for the high dispersion that we find in the high $M_*$ regime. Similarly, the $sM_{\rm dust}$ trend with $M_*$ of the Ri22 simulated galaxies are in good agreement with our observed galaxies (especially if we apply the corrections expected for \citealt{Kroupa2001}). However, a larger difference is found in the low-$M_*$ regime. This could be attributed to higher gas fractions assumed by the models, compared to the ones in the observed sample, as suggested also by their more massive models showing that the dispersion of the $sM_{\rm dust}$ for a specific $M_*$ depends on the gas content of the galaxies. Specifically, for their galaxies with $\log_{10}(M_*/\rm M_{\odot})\simeq10$ they find different $sM_{\rm dust}$ by assuming different gas fraction. Regarding $\log_{10}(sM_{\rm dust})$ versus $\log_{10}$(sSFR), shown in the right panel of Fig.~\ref{fig:models_literature} the models show a larger dispersion than the one suggested by our data. The As13 model with the lowest $\tauup_{\rm SF}$ lies on top of our data, while higher $\tauup_{\rm SF}$ gives higher $sM_{\rm dust}$ than what we find. Similar $sM_{\rm dust}$ is predicted by the Ca09 dwarf galaxy model. On the other hand, the higher mass Ca09 models are in better agreement with our data, following though, a steeper decrease of $sM_{\rm dust}$, as we go at higher masses (lower sSFR), than that found for our sample. The Ga21 models do not agree with our data in the $\log_{10}(sM_{\rm dust})$--$\log_{10}$(sSFR) plane, however we need to stress out that as stated by them this might reflect a simplistic SFH, outflow, and inflow prescription used in their models. For this reason, we present the properties of models with lower outflow rates, at 10 Gyr, as at later ages they give very small sSFR for a constant $sM_{\rm dust}$, and the ones of the model with high outflow rates, at only 4.6 Gyr, since after that age, it runs out of ISM.

\subsection{Our evolutionary models estimated with \texttt{chemevol}}

The comparison of our observed galaxies with the model predictions from the literature indicates that different model assumptions and initial conditions can lead to different results. In the following, we use \texttt{chemevol} to study the dependence of the observed trends on the various parameter models and we investigate how the combined DustPedia-LVL data can help in constraining the properties of the models.

\begin{table}
\centering
\footnotesize
\caption{Dust evolution parameter grid in \texttt{chemevol}.}
\label{tab:grid_chemevol}

\begin{threeparttable}
\begin{tabular}{@{}llr@{}}
\hline\hline
Parameter & Notation & Values \\
\hline
Model start time [Gyr] 
& $t_{\rm start}$ 
& 1, 6, 11 \\

Initial gas mass [M$_\odot$] 
& $M_{\rm gas,\,init}$ 
& 1E9, 5E10, 1E12 \\
\multicolumn{3}{@{}l@{}}{} \\

Reference star 
& 
& fast.sfe: 1E$-$8.5 \\

formation efficiency & SFE$_0$$^{(\dagger)}$ 
& \underline{average.sfe: 1E$-$9.0} \\

[yr$^{-1}$] & 
& late.sfe: 1E$-$9.5 \\

\multicolumn{3}{@{}l@{}}{} \\

Fraction of gas in dense clouds 
& $f_{\rm c}$ 
& 0.03, \underline{0.5} \\

Outflow rates reduction factor 
& $f_{\rm out}$ 
& 0.25, \underline{1.0} \\

Photofragmentation rate & & \\
of large grains & $k_{\rm frag}$$^{(\star)}$ 
& \underline{0.03}, 0.50, 5.00 \\

Grain growth scaling factor 
& $k_{\rm gg,\,cloud}$$^{(\star)}$ 
& 500, 1000, \underline{3829} \\

SN dust yields reduction factor 
& SN$_{\rm red}$ 
& \underline{2}, 20, 80 \\
\hline\hline
\end{tabular}

\begin{tablenotes}[flushleft]
\footnotesize
\textbf{Notes.} Only those parameters that are varied in this work are listed.
Values from the DV21 reference model are underlined.
For the rest of the parameters, not listed here, we selected the values of the
reference model as presented in DV21 (see their Sec.~6).
$^{(\dagger)}$ See Eq.~(\ref{eq:SFE}).
$^{(\star)}$ \texttt{THEMIS} model parameters; see Eq.~(11) of DV21.
\end{tablenotes}
\end{threeparttable}
\end{table}

We already mentioned some details of \texttt{chemevol} above, but we describe here the basic features of the model. The code solves differential equations that account for the secular evolution of the main building blocks of galaxies, i.e. stars, gas, heavy elements and dust, under the assumption that they are perfectly mixed. The ISM is separated into clouds and diffuse ISM. 
Stars form from gas (pristine clouds initially), a portion of which is returned into the interstellar medium (ISM) at the end of their lifetimes. The model follows the stellar evolution as a function of the mass, which determines both the lifespan and the yields of elements and dust. 
The stellar component is strongly dependent on the assumed initial mass function (IMF). Gas is consumed through processes, such as astration and galactic outflows, while it is replenished via stellar feedback and inflow of metal- and dust-free gas. 
Heavy elements are infused in the ISM at the end of the stellar lifetime. A part of them returns into stars through astration and a fraction is lost through outflows. Dust is produced by three main processes, which are the condensation of elements into solid grains occurring in low- and intermediate mass stars ejecta, the condensation in type-II SN ejecta, and the grain growth through accretion from elements in the ISM.

For our investigation, we use as reference the \citet[DV21]{DeVis2021} best-fit model, assuming three initial gas masses ($M_{\rm{gas,~init}}$) aiming to cover the mass range of our combined sample. For each $M_{\rm gas, ~init}$ we assume three start times ($t_{\rm start}$) in order to estimate the properties of both younger and more evolved galaxies. The end time of the models is always fixed to 13.8 Gyr. Consistently with our \texttt{CIGALE} analysis, the best-fit model uses a \citet{Salpeter} IMF, the \texttt{THEMIS} dust model is adopted, while a delayed SFH is assumed. The SFH prescription adopted by DV21 is given by
\begin{equation}\label{eq:SFE}
    {\rm SFE} = {\rm SFE}_0 \left(\frac{M_*}{10^9}\right)^{0.25} \left(1+\exp^{M_*/10M_{\rm gas}}\right)^{-3} (1+z)^{-1},
\end{equation}
where $z$ is the redshift, $M_{\rm gas}$ is the gas mass, and SFE$_0$ is the reference SFE, which is a free parameter. As shown by DV21 (see their Sec. 3.3 for details), from the above empirical prescription, the resulting SFH is consistent with the delayed SFHs observed in other works. We vary the key parameters of the model within their  suggested rate (as listed in Tab.~1 of DV21). The reference model (denoted as ref) and its variations are shown in Fig.~\ref{fig:OurModels12}. The parameters investigated are listed in Tab.~\ref{tab:grid_chemevol}.

\subsection{Qualitative inspection of dust evolution parameters}

\begin{SCfigure*}[1.2][t!]
\centering
\includegraphics[width=0.8\textwidth]{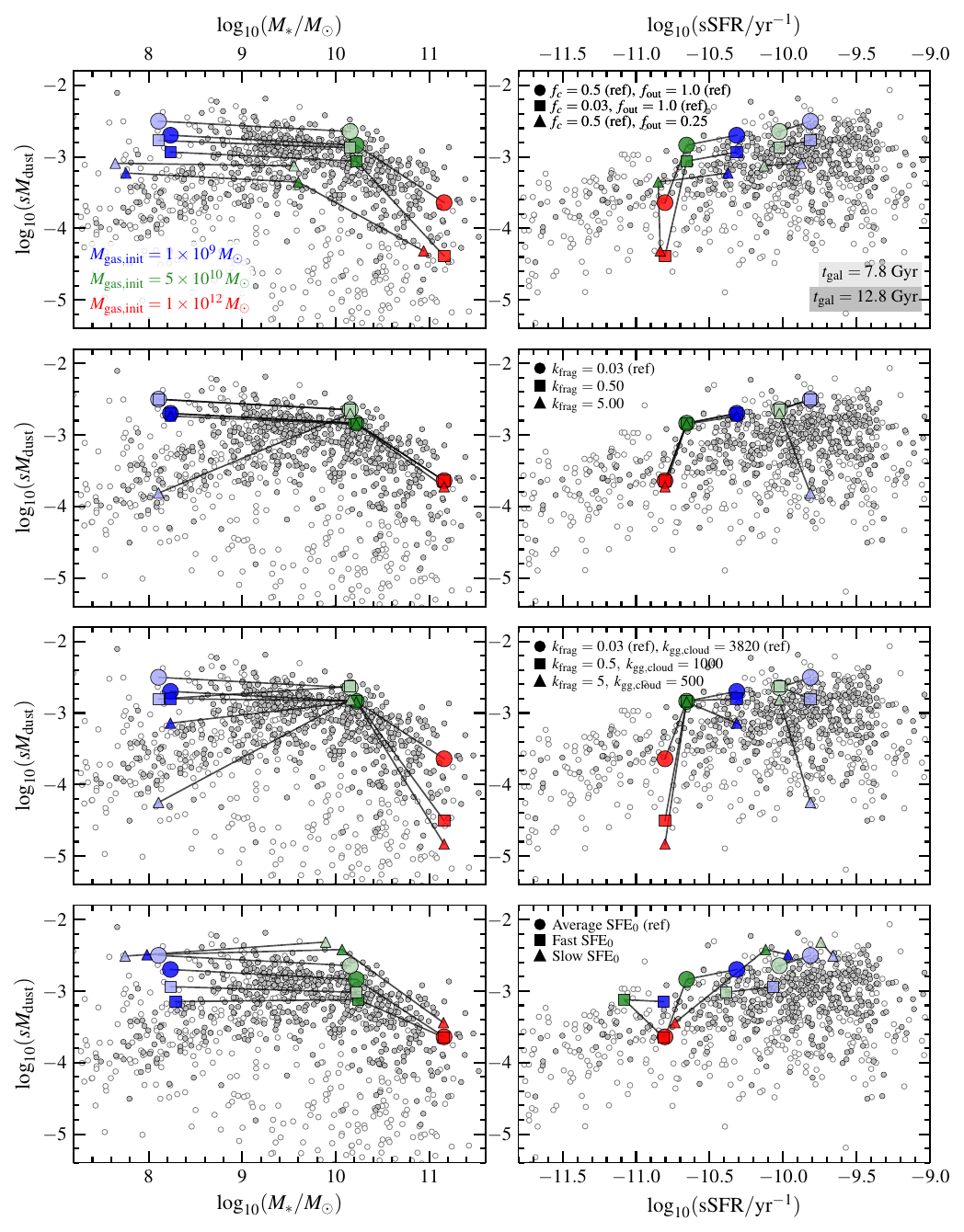}
\caption{$sM_\mathrm{dust}$ as a function of $M_{*}$ (\textit{left panels}) and sSFR (\textit{right panels}) compared to our model predictions using \texttt{chemevol}. The colour coding of data points is the same as in Fig.~\ref{fig:models_literature}. Model estimates are colour coded by $M_{\rm gas,~ init}$, with darker colour corresponding to older galaxies (larger galaxy age, $t_{\rm gal}$). Apart from $M_{\rm gas,~ init}$ and $t_{\rm gal}$, in each row we also vary and indicate with a different symbol (i.e. circle, square, and triangle): the cold gas fraction ($f_{\rm c}$) and the outflow reduction factor ($f_{\rm out}$; \textit{first row}), the \texttt{THEMIS} photofragmentation rate of \mbox{a-C:H/a-C} grains ($k_{\rm frag}$; \textit{second row}), both $k_{\rm frag}$ and the cloud grain growth scaling factor from \texttt{THEMIS} ($k_{\rm gg,~cloud}$; \textit{third row}), and the reference SFE (\textit{fourth row}). Black solid lines correspond to the isochrones.
\label{fig:OurModels12}}
\end{SCfigure*}

In this investigation, for each parameter configuration, our starting point is the assumption of three $M_{\rm{gas,~init}}$ values and two galaxy ages ($t_{\rm gal}$). As can be seen in the top panels of Fig.~\ref{fig:OurModels12} the initial gas budget has the most significant effect on the present day position of galaxies on both the planes studied. A larger assumed $M_{\rm{gas,~init}}$ supplies additional material for galaxies, facilitating the attainment of higher stellar mass. In our models, half of the adopted initial gas mass is assumed to be in the form of inflows. If we vary this fraction (not shown here), we observe exactly the same behaviour as by varying the $M_{\rm{gas,~init}}$. Moreover, younger galaxies (paler colours) with same assumed model parameters systematically have lower $M_*$ (left panel) and higher sSFR (right panel). While in the $\log_{10}(sM_{\rm dust})$--$\log_{10}(M_*)$ diagram the galaxy age leads to variation of the properties along both x- and y-axes, in the $\log_{10}(sM_{\rm dust})$--$\log_{10}$(sSFR) diagram, different galaxy ages correspond mainly to variations along x-axis.

A gas-related parameter (panels of first row) is the fraction of gas that lies in cold dense clouds ($f_{\rm c}$; default value is 0.5; As13). We see in the $\log_{10}(sM_{\rm dust})$--$\log_{10}(M_*)$ plot that when the fraction is reduced to $f_{\rm c}=0.03$ the galaxy ends up having the same $M_*$, but slightly lower $sM_{\rm dust}$. The dependence of the $sM_{\rm dust}$ variation on the $f_{\rm c}$, seems to be more significant at higher stellar masses ($\sim0.4~\rm dex$ lower at $\log_{10}(M_*/\rm M_{\odot})\simeq11$), explaining partially the dispersion observed by our data in that regime. Similarly with the $M_{*}$, in the $\log_{10}(sM_{\rm dust})$--$\log_{10}$(sSFR) diagram, the $f_{\rm c}$ does not affect the sSFR levels. In the same panels we also vary the outflow rates and specifically the \texttt{outflows: reduce} parameter (here $f_{\rm out}$). The outflow rates are defined by \citet{Nelson2019} and $f_{\rm out}$ is not considered as a free parameter in DV21. The architecture of \texttt{chemevol} allows us to vary it though. We stress that $f_{\rm out}$ is a factor that reduces the outflow rate, so by setting it below unity (e.g. 0.25 here), the outflow rate increases. So, when we increase the outflow rate we get galaxies with lower $M_*$ (by 0.1 to 0.3 dex, depending on $M_{\rm{gas,~init}}$), as expected, and lower $sM_{\rm dust}$ (by $\sim0.3~\rm dex$). Regarding the $\log_{10}(sM_{\rm dust})$--$\log_{10}$(sSFR) correlation, lower mass galaxies with higher outflow rate have not only lower $sM_{\rm dust}$, but also slightly reduced sSFR. However, in the most massive galaxies, the outflows affect more significantly the $sM_{\rm dust}$, with their impact on the sSFR being negligible. Generally, $f_{\rm out}$ affects all the physical properties examined here more significantly than $f_{\rm c}$, as it leads to the reduction of the ISM materials available for stellar and dust formation. A similar behaviour is observed by varying the outflow recycling time factor (not shown here). Longer recycling times lead to lower $sM_{\rm dust}$ and $M_*$. Among all the aforementioned models computed by varying gas-related parameters, the model for $f_{\rm out}=0.25$ in intermediate and low mass galaxies is closer to the average values of our data in the $\log_{10}(sM_{\rm dust})$--$\log_{10}$(sSFR) diagram. However, the gas-related parameters fail to reproduce low-mass galaxies with low $sM_{\rm dust}$, and high sSFR, as well as high-mass galaxies with low $sM_{\rm dust}$ and sSFR, in each of the plots, respectively. This is apparent by the isochrones; the curves that cross the data points of galaxies with same age and parameter configuration, but different mass (see solid black lines in Fig.~\ref{fig:OurModels12}). The observed $\log_{10}(sM_{\rm dust})$--$\log_{10}(M_*)$ trend of our data for $\log_{10}(M_{*}/ {\rm M_\odot})\gtrsim9$ is reproduced by the isochrones; however, by varying gas-related parameters, we find that the models systematically overestimate the $sM_{\rm dust}$ compared to the observed values for lower-mass galaxies. In the case of the $\log_{10}(sM_{\rm dust})$--$\log_{10}(\mathrm{sSFR})$ relation, the corresponding isochrones of both young and old low-mass galaxies follow the observed trend, and the massive galaxies deviate from it.

Apart from the gas-related properties, we examined also the effects of varying dust-related parameters. In the second row of panels we vary the {\tt THEMIS} photofragmentation rate of large a-C:H/a-C grains ($k_{\rm frag}$). We find that $k_{\rm frag}$ affects the youngest (7.8 Gyr old) and lowest stellar-mass galaxies. In such objects, the highest $k_{\rm frag}$= 5.00, leads to lower dust content, reducing its $sM_{\rm dust}$ by more than an order of magnitude. Young low-mass galaxies with high photofragmentation rates are the only ones that occupy the same area defined by our low-mass Sm-Irr galaxies, following the increasing trend in the $\log_{10}(sM_{\rm dust})$--$\log_{10}(M_*)$ diagram, and lie in the low $sM_{\rm dust}$ - high sSFR regime in the $\log_{10}(sM_{\rm dust})$--$\log_{10}(\mathrm{sSFR})$ diagram. Additionally, if we decrease the cloud grain-growth scaling factor $k_{\rm gg,~cloud}$ together with the increase of $k_{\rm frag}$, the dust content varies more significantly, and a difference in $sM_{\rm dust}$ is found, also for the older low-mass, as well ass for the older high-mass galaxies (fourth row of panels in Fig.~\ref{fig:OurModels12}). Intermediate mass galaxies are not significantly affected by either of the two aforementioned parameters. Another dust-related parameter that we explored, is the SN$_{\rm red}$, a factor that reduces the amount of dust that is formed in SN and reaches the ISM \citep[for example, because it is destroyed by the reverse shock within the remnant; see, e.g.][]{Bocchio2016}. However, we do not plot it because SN$_{\rm red}$ only affects the $sM_{\rm dust}$ at the very early stages of the galaxies evolution, when SNe dominate dust production. At later stages, as the ones we investigate, where grain growth in a higher metallicity ISM is dominant, the properties are not affected by this factor.

Finally, as mentioned already, the prescription of the adopted delayed SFH in Eq.~(3) allows us to vary the SFE by assuming different reference SFE$_{0}$ (see Tab.~\ref{tab:grid_chemevol}). Hence, in the bottom panels of Fig.~\ref{fig:OurModels12} we compute the models assuming various SFEs. We find that less massive galaxies are more sensitive in changes in the assumed SFE. As suggested by the models, galaxies with fast SFE attain somewhat higher $M_*$ (up to 0.2 dex), and lower $sM_{\rm dust}$ and sSFR (up to 0.5 dex and 0.7 dex, respectively), compared to their counterparts of the same age and an average SFE. The assumption of a slow SFE has the opposite effect. Massive galaxies [$\log_{10}(M_*/\rm M_{\odot})\approx11$] are less affected by the various SFEs assumed.

Overall, the comparison between our observed galaxies properties and the theoretical models indicates that the observed $\log_{10}(sM_{\rm dust})$--$\log_{10}(M_{*})$ trend can be interpreted mainly by varying the initial gas mass in the models, while the $\log_{10}(sM_{\rm dust})$--$\log_{10}(\rm{sSFR})$ trend is mainly explained by varying galaxy ages. In general, while gas-related parameters affect similarly the galaxies, independent their age and mass, low-mass galaxies especially the younger ones are more sensitive in differences of the dust-related model parameters. Intermediate-mass galaxies are the least sensitive to different dust related assumptions and the properties of high-mass galaxies are not affected by the various assumed SFEs. The properties of our observed galaxies in the $\log_{10}(M_*/\rm M_{\odot})<9$ and $\log_{10}({\rm sSFR/yr}^{-1})>-10$ regimes, can be only reproduced by modelled galaxies with $t_{\rm gal} = 7.8$ Gyr and high $k_{\rm frag}$, low $k_{\rm gg,~cloud}$. The model cannot reproduce the observed $sM_{\rm dust}$ for the latter subset, if only gas-related properties or the SFE are varied. Taking into account the fact that in \citet{DeVis2017_464} this regime is occupied only by their gas-rich low-stellar mass galaxies (\textsc{HiGH}-low subset), only variations in the dust-related properties can explain the trends for these objects (not only the initial low gas mass budget). Moreover, the early-type galaxy (Es and S0) could not be reproduced by any of the model configurations that we assumed. In general, the current exercise also highlights the diversity of galaxy evolution paths and the need for sophisticated, galaxy-by-galaxy, modelling approaches, as also illustrated by \citet{Calura2023} with a model tailored to M74 (NGC0628, a DustPedia galaxy). The combined DustPedia-LVL sample can help in the direction of constraining such models. 

\section{Summary and conclusions}\label{sec:sum}

We merged the DustPedia and LVL samples, obtaining a database of 1011 local galaxies. We estimated the physical properties of these galaxies through homogeneous SED fitting. We have shown that the combined sample includes galaxies of all morphological types, spanning a wide range of masses and star-formation activity, well suited to study the evolution of the dust content in the local Universe. Specifically, we studied the correlation of the $sM_{\rm dust}$ as a function of $M_*$ and sSFR. Our main results are:

\begin{itemize}

\item the trend of $\log_{10}(sM_{\rm dust})$ vs $\log_{10}(M_*)$ is not monotonic. A positive correlation is found for galaxies with $\log(M_*/\rm M_\odot)<9.5$, and a negative one for higher masses.

\item The peak of the $sM_{\rm dust}$ in local galaxies, and the change in the slope of the correlation, occurs at $\log_{10}(M_*/{\rm M}_\odot) = 9.5$.

\item The sSFR of high-mass galaxies establishes a strong linear correlation with $sM_{\rm dust}$. However, a large fraction of E and S0 sources lies below the linear trend, at the low-sSFR regime.

\item Low-mass spirals (Sa-Sdm), irregulars and dwarf galaxies (Sm-Irr) lie in the high-sSFR regime, with an increased dispersion below the linear trend found for high-mass galaxies.

\item Grouping the galaxies in bins of stellar mass, we find that $sM_{\rm dust}$ increases for low-mass galaxies that have high sSFR, reaching a peak $sM_{\rm dust}$ at $\log_{10}(\mathrm{sSFR}/{\rm yr}^{-1})\simeq-9.9$. A similar behaviour is found by grouping the galaxies in bins of morphological type, with the peak occurring at $0.1~{\rm dex}$ higher $\log_{10}(\mathrm{sSFR})$.

\end{itemize}

We compared our results with evolutionary models provided in the literature and models that we computed using the one-zone dust evolution model of \citet{DeVis2021}, by varying specific free parameters of their reference model. We find that:

\begin{itemize}

\item the observed trends in the $\log_{10}(sM_{\rm dust})$--$\log_{10}(M_{*})$, and the $\log_{10}(sM_{\rm dust})$--$\log_{10}$(sSFR) planes can be interpreted mainly as the result of different initial gas mass budget and different galaxy ages, respectively.

\item While gas-related parameters influence the positions of galaxies of all masses and ages in the examined diagrams, massive old galaxies are not sensitive in SFE variations. Dust-related parameters affect more significantly low-mass galaxies (especially the young ones) and the more massive-old galaxies. The properties of intermediate mass galaxies are not affected by the different assumptions of dust-related parameters.

\item The observed properties and trends of low-mass and highly star-forming Sm-Irr galaxies can be reproduced by the models if we assume a young galaxy age, a high efficiency for the photofragmentation rate of large grains, and/or low grain-growth scaling factor in star-forming clouds.

\end{itemize}

The combination of DustPedia and LVL constitutes an ideal sample for studying dust evolution in nearby galaxies and could be used to constrain chemical evolution models. While a significant fraction of DustPedia galaxies has available observations for atomic, molecular gas, and metallicity \citep{DeVis2019, Casasola2020, Salvestrini2025}, allowing to set further constraints  to evolutionary models \citep[see, e.g.][]{Galliano2021,DeVis2021}, this is not the case for LVL. Future observations (or, possibly, 
literature compilations) for LVL galaxies, including \textsc{Hi} 21~cm, $^{12}$CO emission lines, optical spectroscopy, and dust emission up to the submm, are needed to confirm and better interpret the scaling relations found in the current study. In particular, observations with the proposed PRobe far-Infrared Mission for Astrophysics \citep[PRIMA\footnote{\url{https://prima.ipac.caltech.edu/}};][]{Glenn2025} will help to constrain better their total dust budget and properties \citep[see e.g.][]{Casasola2025prim, Traiana2025prim, Galliano2025prim}.


\begin{acknowledgements}
We thank the anonymous referee, whose constructive comments and suggestions helped to clarify and improve the content of this study.
We thank Daniela Calzetti, for drawing our attention to the "diversity" of the DustPedia and LVL results, and M\'ed\'eric Boquien, for constructive discussions about \texttt{CIGALE}.
EDP, SB, and EC, acknowledge financial support from INAF-Mini Grant 2024 program “Dust emission and optical extinction as gas tracers in star forming galaxies”.
VC, SB, FC, FP, and VT acknowledge financial support from INAF-Mini Grant 2024 program “DustPedia meets Metal-THINGS: Dust-METAL”.
EC acknowledges financial support from INAF-Mini Grant 2023 program “SHAPES”.
VC, SB, FC, and FP acknowledge financial support from INAF-Mini Grant 2022 program “Face-to-Face with the Local Universe: ISM’s Empowerment (LOCAL)”.
FG acknowledges support by the French National Research Agency under the contracts WIDENING (ANR-23-ESDIR-0004) and REDEEMING (ANR-24-CE31-2530), as well as by the Actions Thématiques “Physique et Chimie du Milieu Interstellaire” (PCMI) of CNRS/INSU, with INC and INP, and “Cosmologie et Galaxies” (ATCG) of CNRS/INSU, with INP and IN2P3, both programs being co-funded by CEA and CNES.
DustPedia is a collaborative focused research project supported by the European Union under the Seventh Framework Programme (2007–2013) call (proposal no. 606824). The participating institutions are: Cardiff University, UK; National Observatory of Athens, Greece; Ghent University, Belgium; Université Paris-Sud, France; National Institute for Astrophysics, Italy and CEA (Paris), France.
This research made use of \texttt{Astropy}: a community-developed core Python package and an ecosystem of tools and resources for astronomy \citep[][\url{http://www.astropy.org}]{astropy:2013, astropy:2018, astropy:2022}, \texttt{matplotlib}, a Python library for publication quality graphics \citep{Hunter2007}, \texttt{NumPy} \citep{Harris2020}, \texttt{SciPy} \citep{SciPy-NMeth}.

\end{acknowledgements}

\bibliographystyle{aa}
\bibliography{References}

\appendix

\section{Tests on SED fitting}
\label{app:tests}

As described in Sect.~\ref{sec:CIGALE}, in this work we fit the SED of LVL galaxies with \texttt{CIGALE} and the same parameter space used for DustPedia galaxies by \citet{Nersesian2019}.
Results from \texttt{CIGALE} fits are also provided by \citet{Dale2023}: they used the same {\tt CIGALE} modules as \citet{Nersesian2019} for  SFH, stellar libraries and dust attenuation, but with different choices for some of the parameters
(see their Table~2 for the full {\tt CIGALE} setup).
Regarding the SFH module, the grid of \citet{Dale2023} has a varying age of the burst or quenching episode, but a single galactic age of 13 Gyr.
For the stellar component, the \citet{Chabrier2003} IMF is used, with BC03 stellar population models for a few metallicity values, from sub- to super-solar. In the dust attenuation module, a wider and denser range is adopted for 
the slope of the power law modifier to the Calzetti law \citep{Leitherer2002}.
For the dust emission module, \citet{Dale2023} used the classical \citet{DraineLi2007} emission templates, based on a mixture of silicate and carbonaceous grains (graphite + PAHs), which were updated by \citet[][hereafter DL14]{Draine2014}.

\begin{figure*}[ht!]
\centering
\includegraphics[width=\textwidth]{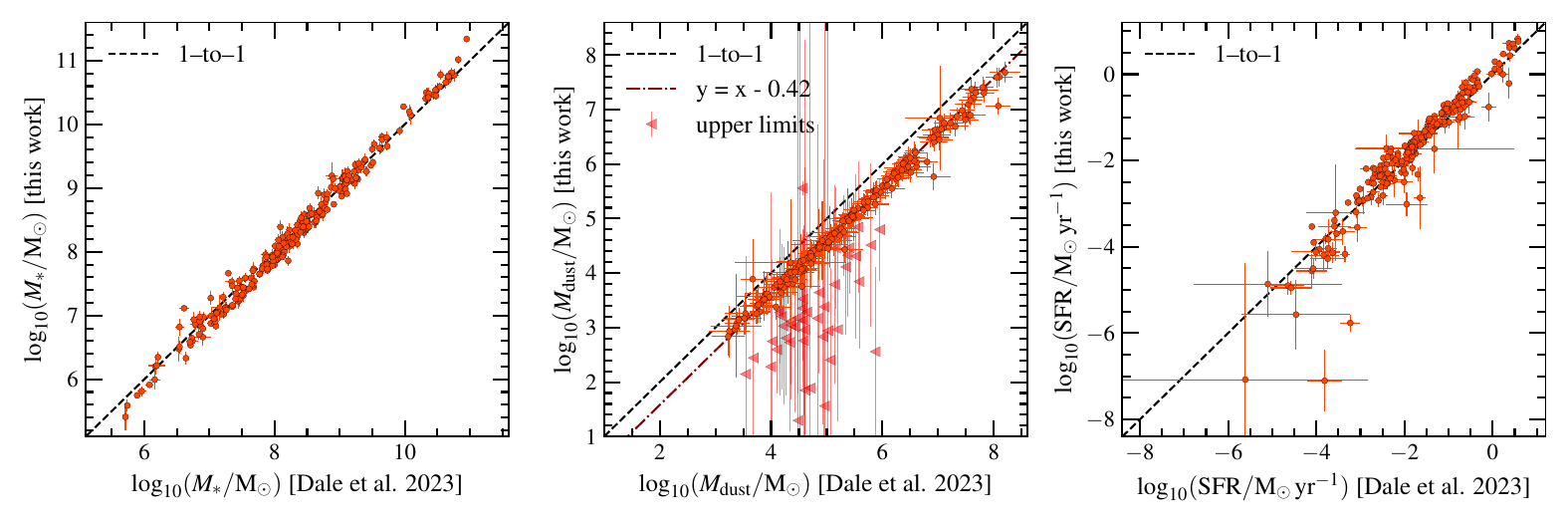}
\caption{Comparison between the physical properties of LVL galaxies, derived using the \texttt{CIGALE} parameter space used in this work \citep[introduced in][]{Nersesian2019} and those of \citet{Dale2023}. Stellar mass, dust mass and SFR are shown in the left, middle and right panel, respectively. A black-dashed line indicates the one-to-one relation. In the middle panel, a dark-red dash-dotted line shows the y= x - 0.42 relation; for 50 objects, \citet{Dale2023} only provide upper limits for $M_\mathrm{dust}$, which are indicated with a different symbol.
\label{app:fig-comp-Ners-Dale}}
\end{figure*}

\begin{figure*}[ht!]
\centering
\includegraphics[width=\textwidth]{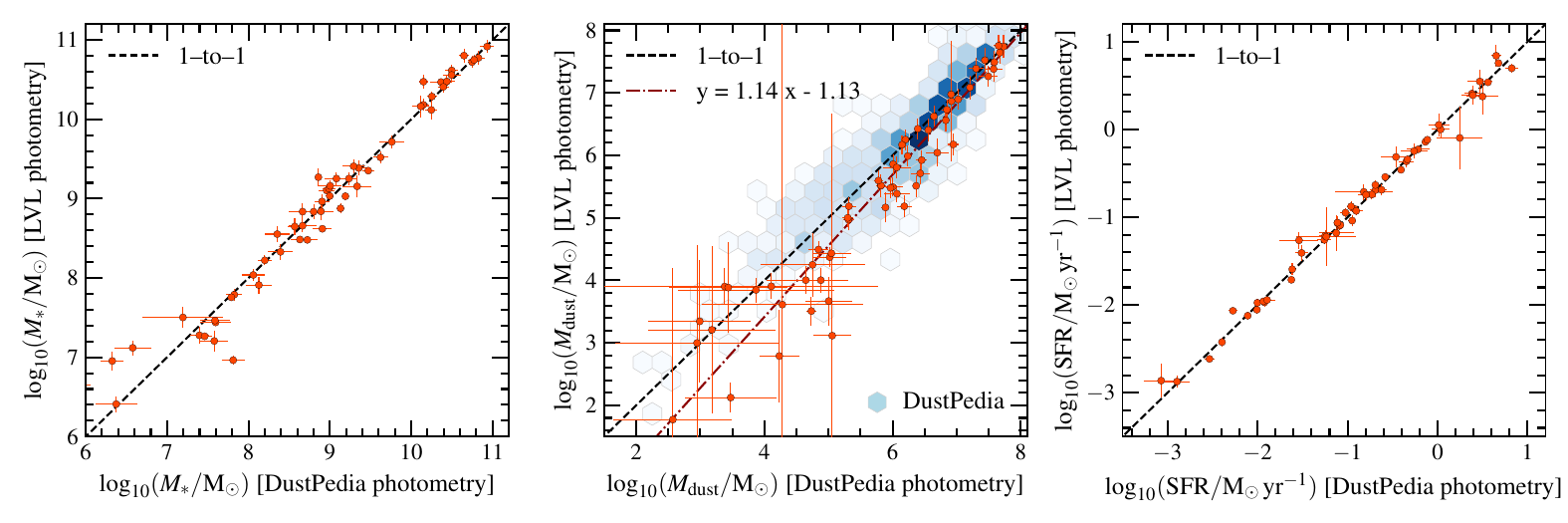}
\caption{Comparison between the physical properties of galaxies in common between the DustPedia and LVL samples, derived from DustPedia photometry (x-axis) and LVL photometry (y-axis) and using the \texttt{CIGALE} parameter space adopted in this work. Stellar mass, dust mass and SFR are shown in the \textit{left}, \textit{middle} and \textit{right panel}, respectively. A solid black line indicates the one-to-one relation.
For both datasets we used the DustPedia distances \citep{Clark2018}. In the \textit{middle panel}, the linear fit to the data is shown with a dark-red dashed line. Also, a density plot indicates the corresponding correlation for all DustPedia galaxies, where in the y-axis we use the dust masses derived by restricting the photometry to the wavelength range of LVL  (i.e. up to 160 $\muup$m), while in the x-axis we use the results derived from the full DustPedia wavelength coverage.
\label{app:fig-comp-LVL-DP}}
\end{figure*}

In Fig.~\ref{app:fig-comp-Ners-Dale} we provide a comparison of the values for $M_*$ (left panel), $M_\mathrm{dust}$ (middle panel) and SFR (right panel) we derived for all the LVL galaxies (including those overlapping with DustPedia) using the \citet{Nersesian2019} parametrisation, and those derived by \citet{Dale2023}. The major difference between the two estimates is for $M_\mathrm{dust}$: the \citet{Dale2023} estimates - based on DL14 - are systematically higher than those we derive with the \texttt{THEMIS} dust model, by $\sim$0.42~dex. This is the same offset 
estimated by \citet{Nersesian2019}, resulting
from the lower emissivity index, $\beta$, and higher fixed absolute opacity, $\kappa(\lambda_0)$, values of \texttt{THEMIS}, that make it more emissive than DL14. In fact, it is well known that the emissivities of the \citet{DraineLi2007} and DL14 templates are biased to low values \citep[for a discussion, see][and references therein]{Galliano2018}; they are also unable to reproduce the common benchmark for dust model, i.e. the high-latitude dust emissivity in the Milky Way \citep{Bianchi2019}. 

We would have expected significative differences also in the estimates of $M_*$ and SFR, due to the different IMF choice in \citet{Nersesian2019} and 
\citet{Dale2023} - which instead are not seen in Fig.~\ref{app:fig-comp-Ners-Dale}. Indeed, if we run our fits by assuming the \citet{Chabrier2003} IMF \citep[adopted by][]{Dale2023} and leaving all other parameters unaltered, $M_*$ is found to be systematically lower than with the \citet{Salpeter} IMF used by \citet{Nersesian2019}, by 0.25 dex; and SFR by 0.2 dex (similar values can be found in the literature, see, e.g., \citealt{2014ARA&A..52..415M,Bernardi2018}). Yet, the different choice of other parameters in the two works also affects the result. In particular, we verified that the longer galactic age assumed by \citet{Dale2023} \citep[13~Gyrs vs 2-12~Gyrs in][]{Nersesian2019} conjures with the wider range in the metallicity of the stellar populations \citep[including subsolar values, while][use solar metallicity only]{Nersesian2019} in raising the mass-to-light ratio of the model: eventually, the offset due to the IMF is compensated and both the \citet{Dale2023} and our fits produce similar $M_*$ and SFR. 

The derivation of the physical properties of galaxies from the SED does not only depend on the selection of the parameter grid in \texttt{CIGALE}, but also on the wavelength coverage. For instance, \citet{Nersesian2021} shows that the shape and the wavelength of the FIR-peak of emission at $\sim$100~$\muup$m, correlate with dust properties, such as the dust temperature, luminosity and mass, and also with the dust-to-stellar mass ratio and the SFR. As already mentioned in Sec.~\ref{sec:sample}, the galaxies in the two samples are well covered from the FUV to the FIR wavelengths. The main difference between them is that the DustPedia galaxies are observed by \textit{Herschel} up to 500~$\muup$m, while the FIR data of LVL galaxies are limited to the MIPS 160 $\muup$m band. In order to check the impact of the different wavelength coverage, we took advantage of the 58 galaxies in common between the two samples,
for which we have two set of \texttt{CIGALE} fits on a common parameter grid: those 
produced in this work using the LVL photometry from \citet{Dale2023}; those obtained from the 
DustPedia photometry by \citet{Nersesian2019}.

A comparison between the two {\tt CIGALE} runs is shown in Fig.~\ref{app:fig-comp-LVL-DP}. 
The change in photometric extent does not affect significantly the estimates for $M_*$ and SFR: the values obtained using the LVL photometry are close to those for the DustPedia photometry, within a small scatter (0.08 and 0.03 dex, respectively). More significant is the scatter for $M_\mathrm{dust}$, a quantity directly derived from the FIR/submm photometry: the scatter between the two estimates is $\approx 0.4$ dex, with an apparent tendency for LVL-based estimates to be lower for lower values of $M_\mathrm{dust}$ (see the linear fit in the middle panel of Fig.~\ref{app:fig-comp-LVL-DP}). In principle, we could use this fit to correct the $M_\mathrm{dust}$ estimates from LVL photometry and align them with the DustPedia results: however, the sample is small and several estimates are uncertain.
As an additional test, we repeated the {\tt CIGALE} fits of the full DustPedia database, but limiting the photometry to data-points with $\lambdaup\le160 \muup$m, to mimic the LVL coverage. The comparison is shown in the central panel of Fig.~\ref{app:fig-comp-LVL-DP}. The fit of the two estimates is along the one-to-one line, with no significant trend. The scatter between the two different DustPedia estimates is similar to that between the estimates with LVL and DustPedia photometry for the 58 galaxies in common. 
Therefore, the reduced FIR-submm spectral coverage of LVL galaxies does not affect significantly the derivation of the physical properties discussed in this work, thereby allowing for the integration and joint examination of DustPedia and LVL. 

\section{Correlation with morphology}
\label{app:morpho}

\begin{figure}[t!]
\centering
\includegraphics[width=0.5\textwidth]{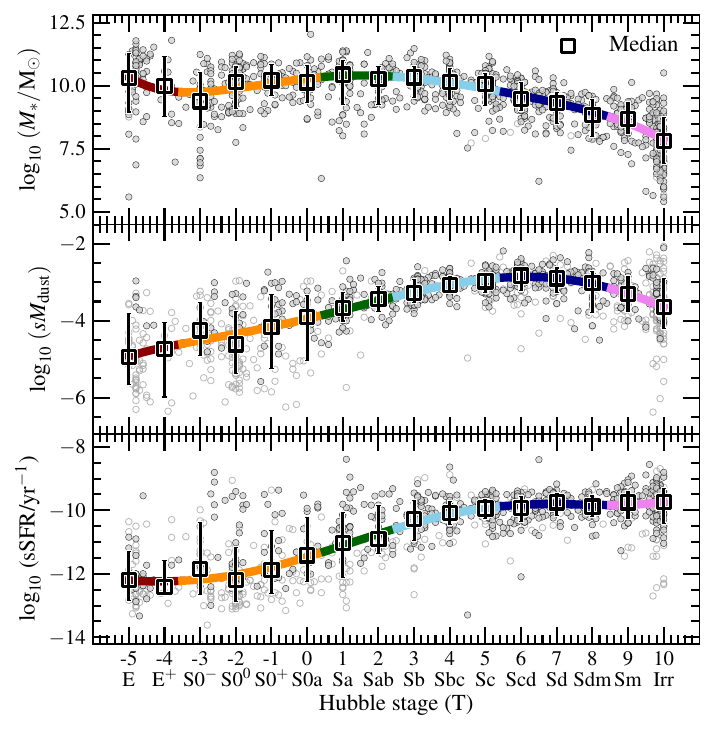}
\caption{Variation of $M_\mathrm{*}$, $sM_\mathrm{dust}$, and sSFR as a function of Hubble stage, from \textit{top} to \textit{bottom panel}, respectively. In each panel grey circles correspond to individual galaxies, black squares are the median values for each morphological bin. Error bars bracket the range between the 16$^\mathrm{th}$ and 84$^\mathrm{th}$ percentiles from the median. The curves, colour-coded with morphology (such as the data-points in Figs.~\ref{fig:R-Mstar-fits} and \ref{fig:R-sSFR-fits}), are the fifth-order polynomial regressions to the median values (see Tab.~\ref{tab:recipes} for the polynomial regression parameters).\\
\label{fig:appPropHT}}
\end{figure}

The dependence of global parameters for DustPedia galaxies, such as $M_*$, $M_\mathrm{dust}$, and SFR, with
morphological type is explored in \citet{Nersesian2019}. In that study, the median values of the parameters, per Hubble stage bin, are computed and then fitted by a $5^\mathrm{th}$ order polynomial, providing the average trend of how the parameters change with morphology (see their Figs. 6, 7, and Table D.1). We apply here the same methodology to the combined sample (DustPedia+LVL) covering a wider range of stellar masses. In Fig.~\ref{fig:appPropHT} we plot the polynomial fits for the physical properties examined in the current work. As can be seen in the top and middle panels,  $M_\mathrm{*}$ and $sM_\mathrm{dust}$  vary on average by two orders of magnitude across different morphologies, while the sSFR variation is $\sim$2.5 orders of magnitude. The maximum of the $M_\mathrm{*}$ is for the E galaxies, with another local maximum at $T\approx1$ (Sa galaxies). In the $sM_\mathrm{dust}$ the maximum is found at $T\approx6$ (Scd). The maximum sSFR  is found for Irr galaxies, and in general late-type galaxies with $5 \lesssim T \lesssim 10$ have two orders of magnitude higher sSFR than earlier-type galaxies with $-5 \lesssim T \lesssim -1$. The estimated values of the coefficients of the polynomial regression of each property are provided in Tab.~\ref{tab:recipes}. These recipes allowed us estimate the values of $M_\mathrm{*}$, $sM_\mathrm{dust}$ and sSFR for a galaxy of a given Hubble stage ($T$). We thus use them to inspect the variation of the correlations investigated [i.e. $\log_{10}(sM_\mathrm{dust}$) -- $\log_{10}(M_\mathrm{*}/\mathrm{M}_\odot)$, Fig.~\ref{fig:R-Mstar-fits}; $\log_{10}(sM_\mathrm{dust}$) -- $\log_{10}(\mathrm{sSFR}/\mathrm{yr}^{-1})$, Fig.~\ref{fig:R-sSFR-fits}] as a function of the morphological stage (see Sec.~\ref{sec:RatioEvolution}).

\begin{figure}[t!]
\centering
\includegraphics[width=0.5\textwidth]{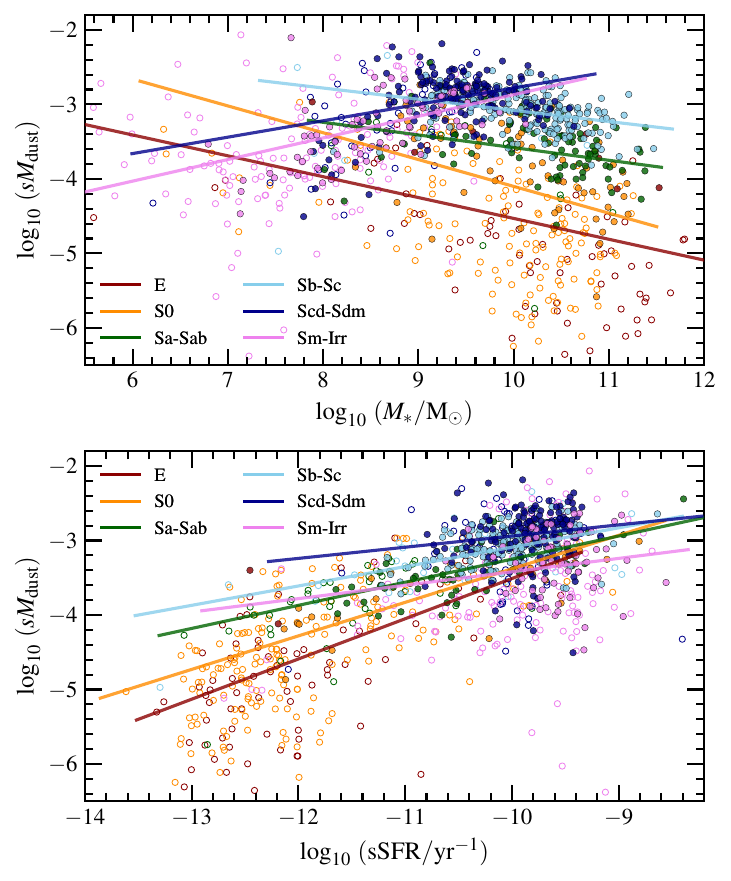}
\caption{$sM_\mathrm{dust}$ as a function of $M_{*}$ (\textit{top panel}) and sSFR (\textit{bottom panel}). Colour-coding is same as in right panels of Figs.~\ref{fig:R-Mstar-fits} and \ref{fig:R-sSFR-fits}. The best linear fits to each morphological bin is shown with the corresponding colour (see the fitting parameters in Tab.~\ref{tab:fit-params}).
\label{app:fig-HubbleFits}}
\end{figure}

The dependence of the correlations on morphology are also traced by performing linear fitting to the galaxies grouped in morphological bins. In Fig.~\ref{app:fig-HubbleFits} we plot the $\log_{10}(sM_\mathrm{dust}$) -- $\log_{10}(M_\mathrm{*}/\mathrm{M}_\odot)$ and $\log_{10}(sM_\mathrm{dust}$) -- $\log_{10}(\mathrm{sSFR}/\mathrm{yr}^{-1})$ relations, with data points colour-coded according to Hubble stage, and with the best-fit models for each morphological bin overplotted. For both investigated correlations, a dependence on the morphology is found. Scd-Sdm and Sm-Irr galaxies show a positive $\log_{10}(sM_\mathrm{dust}$) -- $\log_{10}(M_\mathrm{*}/\mathrm{M}_\odot)$ correlation (steeper for the latter ones), while the other four morphological bins have a negative correlation (steeper for S0s). The $\log_{10}(sM_\mathrm{dust}$) -- $\log_{10}(\mathrm{sSFR}/\mathrm{yr}^{-1})$ correlation is positive for all morphological bins, with the slope getting steeper towards earlier Hubble stages (from Scd-Sdm to E); exception is the Sm-Irr bin which has a steeper relation than the Scd-Sdm galaxies. 

The statistics per morphological bin, along with the fitting parameters and the correlation coefficients, are listed in Tab.~\ref{tab:fit-params}. Despite some morphological types (like E) has large fractions of uncertain measures, the correlations are found to be statistically significant in all cases; following the procedures described in Sect. 5.2 of \citet{Bianchi2018}, we also tested that none of them is driven by the presence of the same quantity on both axis (like, e.g., $M_*$ being used for the x-axis and for the derivation of $sM_\mathrm{dust}$ in the y-axis).
In both correlations, the largest dispersion is found for E galaxies (0.67 and 0.61, respectively), and the smallest for Sb-Sc galaxies (0.21 and 0.18, respectively). According to the correlation coefficients, the strongest linear and monotonic $\log_{10}(sM_\mathrm{dust}$) -- $\log_{10}(M_\mathrm{*}/\mathrm{M}_\odot)$ correlation is found for the ellipticals ($\rho_\mathrm{P}=\rho_\mathrm{S}=-0.52$). It is worth mentioning that if we take into account only the more certain estimates (filled circles), the correlation for Sm-Irr galaxies improves significantly ($\rho_\mathrm{P}=0.61$, $\rho_\mathrm{S}=0.65$). In the case of the $\log_{10}(sM_\mathrm{dust}$) -- $\log_{10}(\mathrm{sSFR}/\mathrm{yr}^{-1})$ relation, apart from the Scd-Sdm and Sm-Irr galaxies, for the rest of the morphological bins the correlation is found to be significant and monotonic ($\rho_\mathrm{P}>0.5$, $\rho_\mathrm{S}>0.4$). Finally, it is worth pointing out that in the combined DustPedia-LVL sample, all Hubble type bins consist of a statistically significant number of galaxies.

\begin{table*}[ht!]
    \caption{Recipes to estimate the integrated physical properties of galaxies given their Hubble stage ($T$).}
    \label{tab:recipes}
    \begin{center}
        \scalebox{0.88}{
        \begin{tabular}{lcccccc}
            \hline 
            \hline 
            \multicolumn{7}{c}{$y = \alpha_0 + \alpha_1 T + \alpha_2 T^2 + \alpha_3 T^3 + \alpha_4 T^4 + \alpha_5 T^5$}\\
            \hline
            $y$ & $\alpha_0$ & $\alpha_1$ & $\alpha_2$ & $\alpha_3$ & $\alpha_4$ & $\alpha_5$ \\
            \hline
            $\log_{10}(M_\mathrm{*}/\mathrm{M}_{\odot})$ 
                & $10.29 \pm 0.10$ 
                & $0.149 \pm 0.041$ 
                & $0.04 \pm 0.01$ 
                & $(-5.75 \pm 1.82)\times10^{-3}$ 
                & $(14.13 \pm 5.74)\times10^{-4}$ 
                & $(-8.04 \pm 4.39)\times10^{-5}$ \\
            $\log_{10}(sM_\mathrm{dust})$ 
                & $-3.95 \pm 0.07$ 
                & $0.22 \pm 0.03$ 
                & $0.01 \pm 0.01$ 
                & $(-1.35 \pm 1.16)\times10^{-3}$ 
                & $(-4.01 \pm 3.65)\times10^{-4}$ 
                & $(2.35 \pm 2.79)\times10^{-5}$ \\
            $\log_{10}(\mathrm{sSFR}/\mathrm{yr}^{-1})$ 
                & $-11.47 \pm 0.08$ 
                & $0.34 \pm 0.033$ 
                & $0.02 \pm 0.014$ 
                & $(-5.65 \pm 1.47)\times10^{-3}$ 
                & $(-2.42 \pm 4.63)\times10^{-4}$ 
                & $(4.20 \pm 3.54)\times10^{-5}$ \\
            \hline 
            \hline 
        \end{tabular}}
    \end{center}
\end{table*}

\begin{table*}[ht!]
    \caption{Best-fit parameters for the correlations examined in Figs.~\ref{fig:R-Mstar-fits} and \ref{fig:R-sSFR-fits}, for different morphological-type bins.}
    \label{tab:fit-params}
    \begin{center}
        \scalebox{0.87}{
        \begin{tabular}{lcc|ccccc|cccccc}
            \hline 
            \hline 
            \multicolumn{3}{c|}{} & \multicolumn{5}{c|}{$\log_{10}(sM_\mathrm{dust})$ -- $\log_{10}(M_\mathrm{*}/\mathrm{M}_{\odot})$} 
            & \multicolumn{5}{c}{$\log_{10}(sM_\mathrm{dust})$ -- $\log_{10}(\mathrm{sSFR})$}\\
            \cline{1-8} \cline{9-13}
            Hubble type & $T$  & N. galaxies & $a$ & $b$ & $\sigma$ & $\rho_\mathrm{P}$ & $\rho_\mathrm{S}$ & $a$ & $b$ & $\sigma$ & $\rho_\mathrm{P}$ & $\rho_\mathrm{S}$ \\
            \hline
            E & [-5.0,-3.5) &  74 & -0.28$^{+0.09}_{-0.09}$ & -1.73$^{+0.97}_{-0.87}$ & 0.67$^{+0.10}_{-0.09}$ & -0.52 & -0.52 & 0.54$^{+0.13}_{-0.14}$ & 1.89$^{+1.62}_{-1.73}$  & 0.61$^{+0.09}_{-0.08}$ & 0.52 & 0.49 \\
            S0 & [-3.5,0.5) & 210 & -0.36$^{+0.07}_{-0.06}$ & -0.50$^{+0.55}_{-0.66}$ & 0.53$^{+0.05}_{-0.03}$ & -0.35 & -0.37 & 0.45$^{+0.04}_{-0.03}$ & 1.12$^{+0.40}_{-0.44}$  & 0.38$^{+0.03}_{-0.03}$ & 0.69 & 0.71 \\
            Sa-Sab & [0.5,2.5) & 95 & -0.17$^{+0.05}_{-0.04}$ & -1.88$^{+0.51}_{-0.49}$ & 0.33$^{+0.03}_{-0.03}$ & -0.26 & -0.36 & 0.31$^{+0.03}_{-0.03}$ & -0.15$^{+0.32}_{-0.34}$ & 0.21$^{+0.02}_{-0.02}$ & 0.71 & 0.70 \\
            Sb-Sc & [2.5,5.5) & 207 & -0.15$^{+0.03}_{-0.03}$ & -1.58$^{+0.29}_{-0.33}$ & 0.21$^{+0.01}_{-0.01}$ & -0.08 & -0.32 & 0.26$^{+0.03}_{-0.03}$ & -0.49$^{+0.33}_{-0.40}$ & 0.18$^{+0.01}_{-0.02}$ & 0.54 & 0.41 \\
            Scd-Sdm & [5.5,8.5) & 205 &  0.22$^{+0.04}_{-0.04}$ & -4.98$^{+0.39}_{-0.35}$ & 0.31$^{+0.02}_{-0.02}$ & 0.47 & 0.26 & 0.15$^{+0.07}_{-0.07}$ & -1.44$^{+0.70}_{-0.78}$ & 0.33$^{+0.02}_{-0.01}$ & 0.16 & 0.19 \\
            Sm-Irr & [8.5,10.0] & 220 &  0.29$^{+0.04}_{-0.05}$ & -5.77$^{+0.41}_{-0.38}$ & 0.43$^{+0.03}_{-0.03}$ & 0.33 & 0.45 & 0.18$^{+0.08}_{-0.10}$ & -1.62$^{+0.78}_{-0.98}$ & 0.48$^{+0.03}_{-0.03}$ & 0.21 & 0.23 \\
            \hline 
            \hline 
        \end{tabular}}
    \end{center}
        \smallskip \phantom{}\\
        \footnotesize
        \noindent
        \textbf{Notes.} Slope $a$, intercept $b$, dispersion $\sigma$, Pearson's correlation coefficient $\rho_\mathrm{P}$, Spearman's correlation coefficient $\rho_\mathrm{S}$.
\end{table*}

\section{Comparison with \citet{DeLooze2020}}
\label{app:d20}

\begin{figure*}[ht!]
\centering
\includegraphics[width=\textwidth]{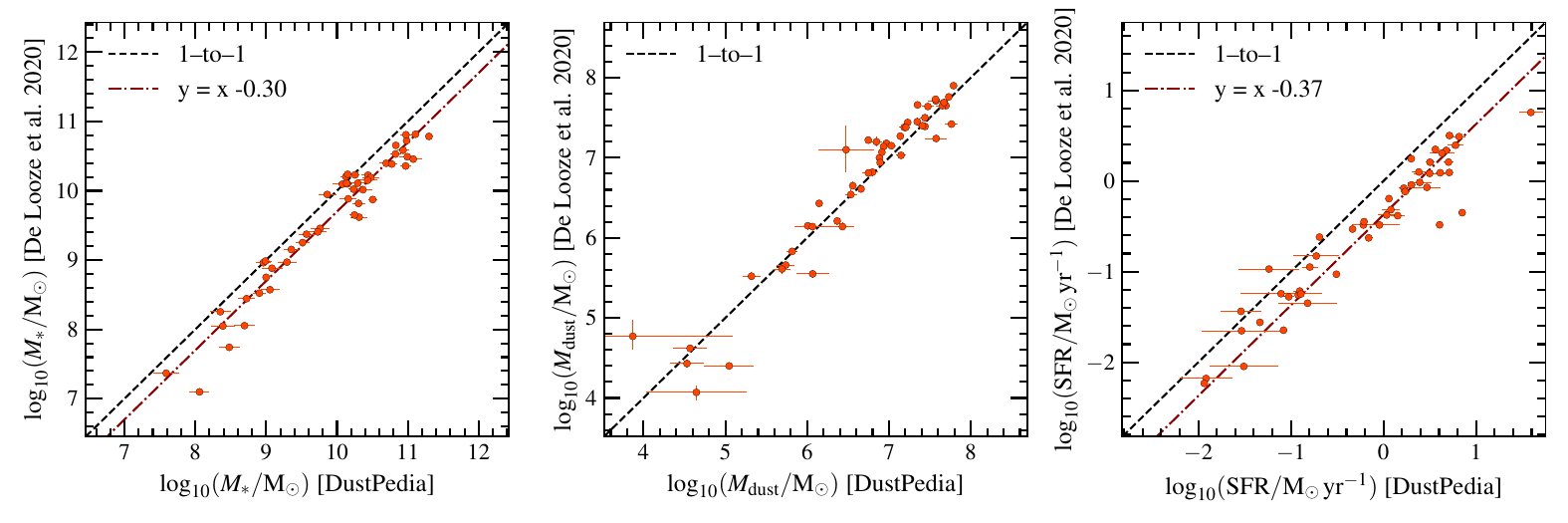}
\caption{Comparison between the physical properties of KINGFISH galaxies as used in the analysis of \citet{DeLooze2020} and those from DustPedia (this work). $M_*$, $M_\mathrm{dust}$ and SFR are shown in the \textit{left}, \textit{middle} and \textit{right} panel, respectively. A solid black line indicates the one-to-one relation and dark-red dash-dotted lines show the systematic offsets (if any). Errors for $M_*$ and SFR for the \citet{DeLooze2020} datapoints are not shown, because they are not available in the table they use \citep[from][]{Hunt2019}.
\label{app:fig-comp-DP-DeLooze}}
\end{figure*}

\citet{DeLooze2020} analysed the combined JINGLE, HRS, KINGFISH, HAPLESS and H\textsc{i}GH sample and found an offset between their $\log_{10}(sM_{\rm dust})$--$\log_{10}(M_*)$ relation and that of \citet{Casasola2020} for DustPedia galaxies. Since both works used the same dust model, \texttt{THEMIS}, the offset was attributed to a selection effect, with the dataset of \citet{DeLooze2020} including more distant (thanks to JINGLE) and dustier galaxies. The analyses of \citet{DeLooze2020} and \citet{Casasola2020} also differed in the way stellar masses and SFRs were derived: while the latter used the {\tt CIGALE} fits of \citet{Nersesian2019} that are also adopted in the current work, the former study by \citet{DeLooze2020} derived $M_{*}$ and SFR using the Multi-wavelength Analysis of Galaxy Physical Properties code \citep[\texttt{MAGPHYS};][]{daCunha2008}.

The large overlap between our datasets \citep[DustPedia-LVL includes about half of the galaxies analysed by][see Sect.~\ref{sec:overlap}]{DeLooze2020} should allow to investigate if the offset is due to selection effects or to different assumptions in the modelling. However, while \citet{DeLooze2020} provide $M_\mathrm{dust}$ for the whole sample, they refer to the literature for $M_{*}$ and SFR; these quantities are available in a tabulated format only for KINGFISH, in \citet{Hunt2019}. The comparison for KINGFISH galaxies is shown in Fig.~\ref{app:fig-comp-DP-DeLooze}. As expected, there is no systematic offset in $M_\mathrm{dust}$ between our results and those of  \citet{DeLooze2020}. Instead, our $M_{*}$ are found to be larger by 0.3 dex, and SFR by 0.37 dex. These systematic offsets are likely due to the differences between the {\tt CIGALE} and {\tt MAGPHYS} parametrization, among which the different choice for the IMF, since {\tt MAGPHYS} uses \citet[][see also Appendix~\ref{app:tests}]{Chabrier2003}.

When showing the results of \citet{DeLooze2020} in Fig.~\ref{fig:R-Mstar-fits} and Fig.~\ref{fig:R-sSFR-fits} we thus applied these offset to their quantities: 0.3 dex for $M_{*}$, -0.3 dex for $sM_\mathrm{dust}$, and -0.07 dex for sSFR. After these corrections, the results of \citet{DeLooze2020} become fully compatible with those of \citet{Casasola2020} and our work.

\end{document}